\documentclass[12pt,journal,final,onecolumn]{IEEEtran} 

\usepackage[pdftex]{graphicx}
\usepackage[cmex10]{amsmath}
\usepackage{amssymb}
\usepackage{amsthm}
\usepackage{amsfonts}
\usepackage{mathtools} 
\usepackage{thmtools} 
\usepackage{bm}
\usepackage{cite}
\usepackage[svgnames]{xcolor} 
\usepackage{pstricks,pst-node,pst-plot,pstricks-add}
\usepackage[tight,footnotesize]{subfigure}
\usepackage[binary-units=true]{siunitx}
\usepackage{hyperref} 

\graphicspath{{figs/}}

\newcommand{\mc}[1]{\mathcal{#1}}

\newcommand{\msf}[1]{\mathsf{#1}}

\newcommand{\defeq}{\mathrel{\triangleq}}
\newcommand{\Pp}{\mathbb{P}}
\newcommand{\E}{\mathbb{E}}
\newcommand{\N}{\mathbb{N}}

\newcommand{\ind}{I}

\DeclarePairedDelimiter\ceil{\lceil}{\rceil}
\DeclarePairedDelimiter\floor{\lfloor}{\rfloor}

\DeclarePairedDelimiter\card{\lvert}{\rvert}

\newcommand{\iid}{i.\@i.\@d.\ }

\DeclareMathOperator{\Bernoulli}{Bernoulli}

\DeclareMathOperator{\head}{head}
\DeclareMathOperator{\tail}{tail}

\newtheorem{lemma}{Lemma}

\newtheorem{theorem}[lemma]{Theorem}

\newtheorem{innercustomtheorem}{Theorem}
{\renewcommand\theinnercustomtheorem{#1}\innercustomtheorem}%
{\endinnercustomtheorem}

\theoremstyle{definition}

\declaretheorem[style=definition,qed=$\lozenge$]{example}
\renewcommand\thmcontinues[1]{Continued}

\newtheoremstyle{myremark}%
{\topsep}{\topsep}{\normalfont}{\parindent}{\itshape}{:}{ }{}

\theoremstyle{myremark}
\newtheorem{remark}{Remark}

\newcommand{\Rs}{R^\star}
\newcommand{\Rf}{R_\msf{FL}}
\newcommand{\Rv}{R_\msf{VL}}
\newcommand{\Rm}{R_\msf{MC}}

\begin{document}

\title{An Information-Theoretic Analysis of Deduplication} 

\author{Urs Niesen%
    \thanks{The author is with Qualcomm's New Jersey Research Center,
    Bridgewater, NJ 08807. Email: urs.niesen@ieee.org}%
    \thanks{This paper was presented in part at IEEE ISIT 2017.}%
}


\maketitle

\begin{abstract}
    Deduplication finds and removes long-range data duplicates. It is
    commonly used in cloud and enterprise server settings and has been
    successfully applied to primary, backup, and archival storage.
    Despite its practical importance as a source-coding technique, its
    analysis from the point of view of information theory is missing.
    This paper provides such an information-theoretic analysis of data
    deduplication. It introduces a new source model adapted to the
    deduplication setting. It formalizes the two standard fixed-length
    and variable-length deduplication schemes, and it introduces a novel
    multi-chunk deduplication scheme. It then provides an analysis of
    these three deduplication variants, emphasizing the importance of
    boundary synchronization between source blocks and deduplication
    chunks. In particular, under fairly mild assumptions, the proposed
    multi-chunk deduplication scheme is shown to be order optimal.
\end{abstract}

\section{Introduction} 
\label{sec:intro}

\subsection{Motivation}
\label{sec:intro_motivation}

Data deduplication is a commonly used technique to reduce storage requirements
in data centers and enterprise servers. It operates by identifying and
removing duplicate blocks of data over long ranges. For example, consider a
corporate logo used in many slide decks of that corporation. The enterprise
storage server, using deduplication, can store only the first occurrence of the
logo and replace subsequent occurrences with pointers to the earlier stored one.

The above example highlights key differences between deduplication and
algorithms used to compress single files.  These latter, by now standard, data
compression approaches include DEFLATE \cite{deutsch96} (based on LZ77
\cite{ziv77} and used in the popular zlib and gzip utilities) and PPM
(prediction by partial match) \cite{cleary84}.  They operate by finding small
amounts of local redundancy. For example, DEFLATE uses a
\SI{32}{\kilo\byte} sliding window and restricts the match length to a maximum
of \SI{258}{\byte} \cite{deutsch96} (although the typical match length is likely
considerably smaller---on the order of a few tens of bytes).  Similarly, PPM
typically uses a context of up to \SI{10}{\byte} \cite{witten91}.  In contrast,
data deduplication finds larger amounts of global redundancy. For example,
\cite{elshimi12} reports finding duplicates on the order of a few up to a hundred
\si{\kilo\byte} over ranges of several hundreds of \si{\giga\byte} up to a few
\si{\tera\byte}. Thus, the main difference between data deduplication and
single-file compression is the scale at which they operate.

To deal with this large scale, deduplication algorithms use an approach called
chunking. In the simplest version, the stream of data (of size up to several
\si{\tera\byte}) is split into chunks of fixed size (say \SI{8}{\kilo\byte}).
The algorithm sequentially processes the stream of chunks. For each chunk, the
algorithm computes a hash value, used as key into a hash table. If the hash
table does not already contain an entry with that key, the algorithm enters the
chunk into the hash table (hash collisions can be avoided by proper
dimensioning of the length of the hash value). The chunk is then deduplicated by
replacing it with its hash value. The hash table and the sequence of chunk
hashes are stored on disk. Since indexing into the hash table can be performed
in constant time, this chunking approach is computationally efficient and can be
performed over large amounts of data.

This fixed-length chunking has the disadvantage that it is susceptible
to shifts of the duplicate data blocks. Returning to the corporate logo
example, if the positions of the logo in the data stream are not aligned
with respect to the chunk boundaries, then duplicates will not be
discovered. To address this issue, most deduplication systems instead
use variable-length chunking, in which the chunk boundaries are defined
by the occurrence in the data stream of short pre-defined anchor
sequences. The chunks now have variable random length.  By choosing the
length of the anchor sequence, the expected length of the chunks can be
controlled. The use of anchor sequences ``resynchronizes'' the
appearance of shifted redundant data blocks, allowing to successfully
deduplicate them.

Deduplication has received significant amounts of attention in the Computer
Systems literature, as surveyed in Section~\ref{sec:intro_related}. It is also
widely used in practice; for example, it is reportedly being used both by
Dropbox \cite{drago12} and by Microsoft Windows Server 2012 \cite{elshimi12}.
Despite its significance, data deduplication seems not to have been studied from
a theoretical point of view.  In particular, an information-theoretic analysis
of its performance limits is missing.

\subsection{Summary of Contributions}
\label{sec:intro_summary}

This paper provides such an information-theoretic analysis of data
deduplication. The main results of this paper are as follows:
\begin{itemize}
    \item It introduces a simple source model, which captures the long-range
        memory and the synchronization issues observed in the data deduplication
        problem.
    \item It formalizes concise versions of the two standard data deduplication
        approaches, one with fixed chunk length and one with variable chunk
        length. It also proposes a third, novel, multi-chunk deduplication scheme.
    \item It analyzes the performance of these three schemes. The fixed-length
        deduplication scheme is shown to be close to optimal when the
        source-block lengths are constant and known a-priori. However, when the
        source-block lengths are variable, fixed-length deduplication is shown
        to be substantially suboptimal. The reason for this suboptimality is
        formally shown to be due to the lack of synchronization between source
        block and deduplication chunk boundaries.
    \item The variable-length deduplication scheme is shown to better handle
        this lack of synchronization. Careful choice of anchor sequence length
        (or, equivalently, expected deduplication-chunk length) is shown to be
        critical for good performance of variable-length deduplication.
    \item Finally, the proposed multi-chunk deduplication scheme is shown to be
        much less sensitive to the expected deduplication chunk length. As a
        consequence, it can better adapt to the source statistics, and has
        order-optimal performance under fairly mild conditions.
\end{itemize}

\subsection{Related Work}
\label{sec:intro_related}

The use of variable-length chunking for the purpose of detecting similar files
in large file systems was proposed in \cite{manber94}. Deduplication based on
this variable-length chunking idea was proposed in \cite{muthitacharoen01}
in the context of a network file system.

The largest gains of data deduplication are achieved when storing different
versions of the same data such as in archival storage \cite{quinlan02,you05} and
backup systems \cite{cox02,zhu08,bhagwat09}. However, data deduplication has
also been successfully applied to primary storage systems
\cite{policroniadis04,elshimi12}.  A further area of application is
virtual machine hosting centers, where data deduplication is used for virtual
machine migration \cite{sapuntzakis02} and for virtual machine disk image
storage \cite{jin09}.

As already mentioned, data deduplication has not yet been investigated from an
information-theoretic point of view. The closest problems in the information
theory literature are compression with unknown alphabets \cite{orlitsky04}, also
known as multi-alphabet source coding \cite{shtarkov95,aberg97}, or the
zero-frequency problem \cite{witten91}. In fact, the large repeated blocks in
the source data can be interpreted as being part of an unknown alphabet that has
to be learned and described by the encoder. 

The related problem of file synchronization has been studied extensively in the
information-theoretic literature
\cite{orlitsky93,orlitsky01a,venkataramanan10,ma11,yazdi14,wang15,rouayheb16}.
The synchronization problem is concerned with duplicates between two versions of
the same file. In contrast, deduplication deals with a large number of
duplicated files or data blocks, and the correspondence between them is not
known a-priori.

\subsection{Organization}
\label{sec:intro_organization}

The remainder of this paper is organized as follows.
Section~\ref{sec:setting} provides a formal definition of the problem
setting, including the source model and the deduplication schemes.
Section~\ref{sec:results} presents the main results.
Section~\ref{sec:conclusion} contains concluding remarks. All proofs
are deferred to appendices.

\section{Problem Setting}
\label{sec:setting}

\subsection{Source Description}
\label{sec:setting_source}

In order to enable an information-theoretic analysis of data
deduplication, we introduce a clean source model. As was pointed out in
Section~\ref{sec:intro_motivation}, the data in deduplication problems
exhibits global long-range dependency: large blocks of data that are
replicated across the entire source sequence. Further, it is unknown
a-priori what and where these repeated blocks are. Rather the
deduplication algorithm, having access to only the source sequence of
zeros and ones, has to discover and then describe any repeated blocks.
Finally, the sizes of the repeats differ from block to block and are
not known a-priori.

We start with a high-level description of the source model. The source
generates a single binary source sequence $\msf{S}$. This source
sequence is the concatenation (without any commas or other delimiters)
of $B$ source blocks, each of which is itself a binary sequence. These
$B$ source blocks are chosen independently and uniformly at random from
a source alphabet. This source alphabet is generated at random and
consists of $A$ randomly chosen binary sequences of variable length. The
goal is to compress the source sequence $\msf{S}$ knowing neither the
source alphabet nor the parsing into the source blocks.

We next provide a formal description of the source model.  We consider a
source \emph{alphabet} $\mc{X}$, of size $\card{\mc{X}}=A$, generated randomly
as follows. Fix a distribution $P_\msf{L}$ over $\N$ with finite mean
$\E(\msf{L})$. Generate $A$ independently and identically distributed
(i.\@i.\@d.\@) random variables $\msf{L}_1, \msf{L}_2, \dots, \msf{L}_A$
from $P_\msf{L}$. We next generate a sequence of binary strings
$\msf{X}_a\in\{0,1\}^{\msf{L}_a}$ for each $a\in\{1, 2, \dots, A\}$:
Starting with $a=1$, choose $\msf{X}_a$ uniformly at random from
$\{0,1\}^{\msf{L}_a}\setminus\{\msf{X}_1, \msf{X}_2, \dots,
\msf{X}_{a-1}\}$.  Thus, each $\msf{X}_a$ is a binary string of length
$\msf{L}_a$, called a \emph{source symbol} in the following. Finally,
the source alphabet $\mc{X}\subset\{0,1\}^*$ (where $\{0,1\}^*$ denotes
the set of all finite-length binary sequences) is defined as the union
of all the source symbols, 
\begin{equation*}
    \mc{X} \defeq \{\msf{X}_a\}_{a=1}^A.
\end{equation*}
Note that $\card{\mc{X}}$ is equal to $A$ since the $\msf{X}_a$ are
drawn without replacement. To simplify some of the derivations later
on, we assume that $\msf{L}$ is tightly concentrated around its mean,
specifically that $\Pp\bigl( \E(\msf{L})/2 \leq \msf{L} \leq
2\E(\msf{L}) \bigr) = 1$. Furthermore, to ensure that the source
alphabet generation is always well defined, we assume that $2 \leq A
\leq 2^{\E(\msf{L})/2-1}$.\footnote{Since $\mathsf{L} \geq
\E(\mathsf{L})/2$, the assumption $A \leq 2^{\E(\mathsf{L})/2}$ would be
sufficient for the source model (in which the source symbols are drawn
without replacement) to be well defined. The additional $-1$ term is
added to simplify the derivations later on.}

From this randomly generated source alphabet $\mc{X}$, we now choose $B$
elements $\msf{Y}_1, \msf{Y}_2, \dots, \msf{Y}_B$ \iid uniformly at
random. Each $\msf{Y}_b$ is referred to as a \emph{source block}. Recall
that each source symbol $\msf{X}_a\in\mc{X}$ is an element of
$\{0,1\}^*$.  Since each source block $\msf{Y}_b$ is equal to a randomly
chosen source symbol, it is therefore also an element of $\{0,1\}^*$,
i.e., a binary sequence of variable length.

Finally, we construct the random \emph{source sequence} $\msf{S}$ as the
concatenation $\msf{Y}_1\msf{Y}_2\cdots \msf{Y}_B$ of $\msf{Y}_1,
\msf{Y}_2, \dots, \msf{Y}_B$.  We emphasize that $\msf{S}$ is an element
of $\{0,1\}^*$, in other words, the boundaries of the source blocks
$\msf{Y}_1, \msf{Y}_2, \dots, \msf{Y}_B$ are \emph{not} preserved.
Denote by $\ell(\msf{S})$ the (random) length of $\msf{S}$.

\begin{example}
    \label{eg:setting}
    Assume the source alphabet is randomly generated to contain the $A =
    5$ source symbols $\mc{X} = \{1, 00, 10, 01100, 001100\}$. From
    this, $B=2$ source blocks are drawn \iid uniformly at random, say
    $\msf{Y}_1 = 10$ and $\msf{Y}_2 = 01100$. The resulting source
    sequence is then $\msf{S} = \msf{Y}_1\msf{Y}_2 = 1001100$. 
\end{example}

Our goal is to compress the source sequence $\msf{S}$, knowing only $P_\msf{L}$,
$A$, and $B$. In particular, the source alphabet $\mc{X}$ and the parsing into
the source blocks $\msf{Y}_1, \msf{Y}_2, \dots, \msf{Y}_B$ are \emph{not} known.
More formally, we are looking for a prefix-free source code for the random
variable $\msf{S}$. We have a complete statistical description of $\msf{S}$ and
(since $\E(\msf{L})$ is finite) its entropy $H(\msf{S})$ is well defined.
The expected rate $\Rs$ of the optimal prefix-free source code for $\msf{S}$
is thus bounded as
\begin{equation*}
    H(\msf{S}) \leq \Rs \leq H(\msf{S})+1.
\end{equation*}

\begin{example}
    \label{eg:setting2}
    We illustrate the order and relative size of the various quantities
    in the problem setting using data from a recent large-scale primary
    (i.e., non backup) data deduplication study~\cite{elshimi12}. The
    length $\ell(\msf{S})$ of the source sequence ranges from several
    hundreds of \si{\giga\byte} to a few \si{\tera\byte}. The expected
    value of the length $\msf{L}_a = \ell(\msf{X}_a)$ of the source
    symbols is a bit harder to quantify (since the notion of source
    symbol itself is a model abstraction), but the experiments
    in~\cite{elshimi12} suggest that reasonable numbers range from a few
    \si{\kilo\byte} to a few \si{\mega\byte}. The number $B$ of source
    blocks is consequently on the order of perhaps $10^5$ to $10^9$.  As
    in our model, the data in~\cite{elshimi12} indicates that duplicates
    are not localized, but occur over the entire source sequence. In
    other words, the source has long-range dependence. The number of
    distinct source symbols $A$ is again hard to precisely quantify.
    Experimental results in~\cite{elshimi12, zhu08} indicate that, depending
    on the scenario, most duplicates occur no more than around
    $100$ times. This suggests that $A$ should be somewhere in the range
    $0.01B$ to $B$. 
\end{example}

There are two key differences between the source model introduced here and the
standard source coding setup. 

First, the standard setting is to consider the source statistics, captured by
$\E(\msf{L})$ and $A$, as fixed and to let the length of the source, captured by
$B$, go to infinity. Instead, as indicated by Example~\ref{eg:setting2}, we are
interested here in the regime when $B$ may be of similar order as $A$. Thus, we
here allow the source parameters $\E(\msf{L})$ and $A$ to grow with $B$. 

Second, given the size of the problem and in particular the long range
over which the source exhibits memory (see again
Example~\ref{eg:setting2}), we are interested in compression schemes
that scale well. As mentioned in Section~\ref{sec:intro}, it is this
scaling requirement of removing large amounts of redundancy (hundreds of
\si{\kilo\byte}) over long ranges (hundreds of \si{\giga\byte}) that
preclude the use of standard compression algorithms such as
LZ77~\cite{ziv77}. Instead, we next describe three deduplication
schemes that do scale well.

\subsection{Deduplication Schemes}
\label{sec:setting_schemes}

We next provide a formal (somewhat stylized) description of the deduplication
approach. There are two types of deduplication schemes that appear frequently in
the literature, fixed-length and variable-length, which are presented first. Then,
we introduce a novel variant of the deduplication approach, termed multi-chunk
deduplication.

For fixed-length deduplication, we fix a chunk length $D$. The source sequence
$\msf{S}$ is parsed into substrings of length $D$ (except for the final
substring that may have length less than $D$). Let $\{\msf{Z}_c\}_{c=1}^\msf{C}$
be this fixed-length parsing of $\msf{S}$ with $\msf{C} \defeq
\ceil{\ell(\msf{S})/D}$.  Each $\msf{Z}_c$ (with $c < \msf{C}$) is an element of
$\{0,1\}^D$ referred to as a \emph{deduplication chunk}. 

The encoding algorithm starts by describing the length $\ell(\msf{S})$
of the source sequence using a prefix-free code for the positive
integers (such as an Elias code~\cite{elias75}). The encoding algorithm
then traverses through the chunks, starting at $c=1$, and constructs a
growing \emph{dictionary} of chunks seen up to $c$. Chunk $c$ is encoded
either as a new dictionary entry or as a pointer into the dictionary at
that point (depending on whether the chunk is new or already in the
dictionary). If chunk $c$ is new, then it is encoded as the bit $1$
followed by the binary string $\msf{Z}_c$ itself. If chunk $c$ is not
new, then it is encoded as the bit $0$ followed by a pointer into the
dictionary.  This fixed-length deduplication scheme is prefix free. Its
expected (with respect to $\msf{S}$) number of encoded bits is denoted
by $\Rf$.

\begin{example}[continues=eg:setting]
    Continuing with Example~\ref{eg:setting}, for $\msf{S} = 1001100$ and with
    $D = 2$, the fixed-length chunks are $\msf{Z}_1 = 10$, $\msf{Z}_2 = 01$,
    $\msf{Z}_3 = 10$, $\msf{Z}_4 = 0$. When the encoding process terminates, the
    chunk dictionary contains the elements $\{10, 01, 0\}$. 

    The encoding of the source sequence length $\ell(\msf{S}) = 7$ is
    $00111$ (using an Elias gamma code). The first two chunks $\msf{Z}_1
    = 10$ and $\msf{Z}_2 = 01$ are not in the dictionary and are
    therefore encoded as $110$ and $101$, respectively. At this point in
    the encoding process, the dictionary is $\{10, 01\}$.  Chunk
    $\msf{Z}_3 = 10$ is equal to the first chunk in the dictionary, and
    is encoded as $00$ (the first $0$ indicating that the chunk is not
    new and the second $0$ indicating the position in the dictionary
    using $\log(2) = 1$ bits). The last chunk $\msf{Z}_4 = 0$ is new and
    is encoded as $10$. The complete encoded source sequence using
    fixed-length deduplication is the concatenation of the various
    encoded chunks and equal to $001111101010010$. 
    
    The decoder decodes this encoded source sequence by first reading
    and decoding the source sequence length $00111$ to $7$. Knowing the
    value of $D=2$, it then traverses the encoded source sequence,
    decoding each chunk and building the dictionary in the process.
\end{example}

\begin{remark}
    Without the initial encoding of $\ell(\msf{S})$, the source code is
    still nonsingular (i.e., no two different $\msf{S}$ have the same
    encoding), and can hence be decoded. However, because of the
    variable-length nature of the source, the code may no longer be
    prefix free. (For example, using fixed-length deduplication with
    $D=2$, the source sequence $0$ would be encoded as $10$ and the
    source sequence $00$ as $100$.) The initial encoding of
    $\ell(\msf{S})$ is therefore necessary to guarantee that the source
    code is prefix free.
\end{remark}

For variable-length deduplication, we fix an anchor sequence, which we take here
to be the all-zero sequence of length $M$ denoted by $0^M$. The source sequence
$\msf{S}$ is then split into a random number $\msf{C}$ of chunks using this
anchor. More formally, the source sequence is parsed as $\msf{Z}_1 \msf{Z}_2
\msf{Z}_3 \cdots \msf{Z}_\msf{C}$, where each chunk $\msf{Z}_c$ (except for
perhaps the last one) contains a single appearance of the sequence $0^M$ at the
end. 

The encoding algorithm again starts by describing the length $\ell(\msf{S})$ of
the source sequence using a prefix-free code for the positive integers.  The
encoding of the sequence itself is also performed using a growing dictionary of
chunks, similar to the fixed-length scheme. If chunk $c$ is new (meaning not yet
in the dictionary), it is encoded as the bit $1$ followed by the binary string
$\msf{Z}_c$ itself. Since the anchor sequence $0^M$ indicates the end of
$\msf{Z}_c$, we do not need to store the length $\ell(\msf{Z}_c)$ explicitly. If
chunk $c$ is not new, then it is encoded, as before, as the bit $0$ followed by
a pointer into the dictionary.  This variable-length deduplication scheme is
also prefix free.  Its expected (with respect to $\msf{S}$) number of encoded
bits is denoted by $\Rv$. 

\begin{example}[continues=eg:setting]
    Continuing again with Example~\ref{eg:setting}, for $\msf{S} = 1001100$ and
    with $M = 2$, the variable-length chunks are $\msf{Z}_1 = 100$, $\msf{Z}_2 =
    1100$. When the encoding process terminates, the chunk dictionary contains the
    elements $\{100, 1100\}$. 

    Once the chunking is completed, the remainder of the encoding
    process for variable-length deduplication is similar to fixed-length
    deduplication. The initial encoding of the source sequence length
    $\ell(\msf{S}) = 7$ is again $00111$. Both chunks $\msf{Z}_1 = 100$
    and $\msf{Z}_2 = 1100$ are not in the dictionary and are therefore
    encoded as $1100$ and $11100$, respectively. The encoded source
    sequence is the concatenation of the various
    encoded chunks and equal to $00111110011100$.

    As for fixed-length deduplication, the decoder decodes this encoded
    source sequence by first reading and decoding the source sequence
    length $00111$ to $7$. Knowing the value of $M=2$, it then traverses
    the encoded source sequence, decoding each chunk and building the
    dictionary in the process. For example, upon seeing the first bit of
    the remaining encoded source sequence $110011100$, the encoder knows
    that the next chunk is new (the initial bit is $1$). It removes the
    $1$ and reads the encoded sequence until the first occurrence of the
    anchor $00$, which results in the string is $100$. This is the
    decoded first chunk, which since it is new is also added to the
    dictionary.
\end{example}

\begin{example}[continues=eg:setting2]
    For a more realistic example, consider again the setting for the primary
    data deduplication study~\cite{elshimi12} from Example~\ref{eg:setting2}.
    The system uses variable-length chunking with expected chunk lengths ranging
    from \SI{4}{\kilo\byte} to \SI{64}{\kilo\byte}. The corresponding anchor
    sequence has length ranging from $12$ to $16$ bits. \cite{elshimi12} finds
    that around \SI{50}{\percent} of chunks appear only once, and that the vast
    majority of chunks have less than $32$ duplicates. It is worth pointing out
    that the number of duplicates may be higher in backup or archival scenarios,
    where deduplication ratios of $20$ to $1$ or higher can be
    achieved~\cite{zhu08}.
\end{example}

We finally describe the novel, multi-chunk deduplication algorithm. We again
split the source sequence into a random number $\msf{C}$ of chunks using the
anchor $0^M$, however, this time we ensure that each chunk has length at least
$2^{M-1}$. More formally, the source sequence is parsed as $\msf{Z}_1 \msf{Z}_2
\msf{Z}_3 \cdots \msf{Z}_\msf{C}$, where each chunk $\msf{Z}_c$ (except for
perhaps the last one) is the shortest string of length at least $2^{M-1}$ 
ending in $0^M$. 

\begin{example}
    \label{eg:setting_mult}
    Multi-chunk deduplication with anchor length $M=4$ parses the source
    sequence $\msf{S} = 1100000000100001000010$ into the chunks $\msf{Z}_1 =
    11000000$, $\msf{Z}_2 = 001000010000$, $\msf{Z}_3 = 10$. 
\end{example}

The encoding algorithm again describes the length $\ell(\msf{S})$ of the source
string using a prefix-free code for the positive integers, followed by a parsing
of $\msf{S}$ using a growing dictionary of chunks. Consider the encoding of
chunk $c$. Assume first that it is new, and consider the sequence $\msf{Z}_{c},
\msf{Z}_{c+1}, \dots$ of chunks.  Let $\msf{V}_c$ be the largest integer such
that $\msf{Z}_{c}, \msf{Z}_{c+1}, \dots, \msf{Z}_{c+\msf{V}_c-1}$ are all new.
These new chunks are then encoded together as the bit $1$, followed by an
encoding of $\msf{V}_c$ using a prefix-free code for the positive integers,
followed by the binary string
$\msf{Z}_c\msf{Z}_{c+1}\cdots\msf{Z}_{c+\msf{V}_c-1}$. Since each chunk
$\msf{Z}_c, \msf{Z}_{c+1}, \dots$ is terminated by the occurrence of the anchor
sequence $0^M$ after position $2^{M-1}-M$, we do not need to store their lengths
explicitly. The encoding process continues with chunk $c+\msf{V}_c$.

Assume next that chunk $c$ is not new, and consider the sequence $\msf{Z}_{c},
\msf{Z}_{c+1}, \dots$ of chunks.  Let $\tilde{c} < c$ be the smallest index
satisfying $\msf{Z}_{\tilde{c}} = \msf{Z}_c$. Such an index $\tilde{c}$ exists
since chunk $c$ is not new; in fact $\tilde{c}$ corresponds to the first time
chunk $\msf{Z}_c$ was seen and hence entered into the dictionary. Consider the
corresponding dictionary entry and the list of subsequent entries. Let
$\msf{W}_c$ be the largest integer such that $\msf{Z}_{\tilde{c}}$,
$\msf{Z}_{\tilde{c}+1}$, \dots, $\msf{Z}_{\tilde{c}+\msf{W}_c-1}$ is equal to
$\msf{Z}_{c}$, $\msf{Z}_{c+1}$, \dots, $\msf{Z}_{c+\msf{W}_c-1}$. Then the
chunks $c$ through $c+\msf{W}_c-1$ are encoded together as the bit $0$, followed
by an encoding of $\msf{W}_c$ using a prefix-free code for the positive
integers, followed by a pointer into the dictionary pointing to chunk
$\msf{Z}_{\tilde{c}}$. Observe that the pointer is to an individual chunk, even
if that chunk was part of a larger group of chunks encoded jointly. The encoding
process continues with chunk $c+\msf{W}_c$.

This multi-chunk deduplication scheme is also prefix free.  Its expected (with
respect to $\msf{S}$) number of encoded bits is denoted by $\Rm$.

\subsection{Performance Metric}
\label{sec:setting_metric}

The standard performance criterion is the redundancy normalized by the
expected length of the source sequence, i.e.,
$(\Rf-\Rs)/\E\ell(\msf{S})$. However, it can be verified that in our
setting $\Rs$ itself may be $o(\E\ell(\msf{S}))$. In these cases, 
\begin{equation*}
    \frac{\Rf-\Rs}{\E\ell(\msf{S})} \leq \frac{\Rf}{\E\ell(\msf{S})} + o(1),
\end{equation*}
which may be small even if $R$ and $R^\star$ are very different.  Thus,
the normalized redundancy may not be a meaningful quantity.  We
therefore instead adopt the ratio $\Rf/\Rs$ (and similar for $\Rv$,
$\Rm$) as our performance metric. This ratio performance metric is
strictly stronger than the standard normalized redundancy
performance metric.\footnote{Indeed, it is easily seen that
\begin{equation*}
    \frac{R-\Rs}{\E\ell(\msf{S})}
    = \frac{\Rs}{\E\ell(\msf{S})} (R/\Rs - 1)
    \leq \bigl(1 + o(1)\bigr) (R/R^\star - 1)
\end{equation*}
as $B\to\infty$.
}

A deduplication scheme with rate $R$ is \emph{order optimal} if $R \leq O(\Rs)$
as $B\to\infty$. It is \emph{asymptotically optimal} if $R \leq (1+o(1))\Rs$ as
$B\to\infty$.

As indicated in Section~\ref{sec:setting_source}, while the standard approach is
to fix the source alphabet parameters, i.e., $\E(\msf{L})$ and $A$, and to
consider the asymptotic behavior as the source length (as measured by $B$)
increases, we are here instead interested in the behavior as the source alphabet
parameters increase together with the source length $B$.

\section{Main Results}
\label{sec:results}

The aim of this paper is to provide an information-theoretic analysis of data
deduplication for the source model defined in Section~\ref{sec:setting_source}.
We start with the fixed-length deduplication scheme as described in
Section~\ref{sec:setting_schemes}. The first result analyzes the performance of
this scheme assuming a constant source-symbol length, i.e., $\Pp(\msf{L} = L) =
1$. As we will see, under this strong assumption, fixed-length
deduplication is close to optimal.

\begin{theorem}
    \label{thm:fixed}
    Consider the source model with $B$ source blocks drawn with replacement from
    the $A$ source symbols of constant length $L$. The performance of the
    fixed-length deduplication scheme with chunk length $D=L$ satisfies
    then
    \begin{equation*}
        1 
        \leq \frac{\Rf}{\Rs}
        \leq 1+7\frac{B+\log(L)}{\min\{A,B\}(L-1) + (B-A)^+\log(A/2)}
    \end{equation*}
    for $B$ large enough.\footnote{All logarithms are to the base two.}
\end{theorem}

The proof of Theorem~\ref{thm:fixed} is reported in Appendix~\ref{sec:proofs_fixed}.
The most interesting regime is when the number of source blocks $B$ is at least
as large as the number of source symbols $A$, in which case the upper bound in
Theorem~\ref{thm:fixed} can be simplified to 
\begin{align*}
    \frac{\Rf}{\Rs}
    & \leq 1+\frac{7B}{(B-A)\log(A/2)}+\frac{7\log(L)}{2(L-1)}.
\end{align*}
Thus, as long as $\omega(1) \leq A \leq (1-\varepsilon)B$ for some $\varepsilon
> 0$ as $B\to\infty$ (which implies that $L = \omega(1)$ as $B\to\infty$ by
assumption), we have that $\Rf \leq (1+o(1))\Rs$ as $B$ grows, showing the
asymptotic optimality of fixed-length deduplication with known and constant
source-symbol lengths

\begin{example}
    \label{eg:running}
    Motivated by Example~\ref{eg:setting2}, consider the source with $A
    = 10^5$ symbols of fixed length $L = 10^6$ bits and with $B = 10^6$ source
    blocks. Theorem~\ref{thm:fixed} shows then that fixed-length deduplication
    with chunk length $D=L$ has performance satisfying
    \begin{equation*}
       \Rs \leq \Rf \leq 1.00007\Rs,
    \end{equation*}
    i.e., very close to optimal.
\end{example}

Unfortunately, the asymptotic optimality of fixed-length deduplication relies
crucially on the assumption of fixed and known source-symbol length.  While this
assumption may be reasonable in certain scenarios (such as for virtual machine disk
image deduplication \cite{jin09}), it is usually not valid. As soon as this
assumption is relaxed, fixed-length deduplication can be substantially
suboptimal, as the next example shows.

\begin{example}
    \label{eg:sync}
    Consider the scenario with $A=2$ source symbols. 
    To start with, assume the source-symbol length is
    constant, $\msf{L}=L=B/3$. By Theorem~\ref{thm:fixed}, fixed-length
    deduplication with chunk length $D=L$ is then within a constant factor of
    optimal as $B\to\infty$.

    On the other hand, assume next that the symbol-length distribution
    $P_\msf{L}$ assigns equal mass to the values $L$ and $L+1$ with $L=B/3$ as
    before. Appendix~\ref{sec:proofs_sync} then shows that fixed-length
    deduplication with chunk length $D=L$ has rate satisfying
    \begin{equation*}
        \frac{\Rf}{\Rs} \geq \Omega(B)
    \end{equation*}
    as $B\to\infty$. In other words, even with only two source symbols, the
    fixed-length deduplication scheme can be substantially suboptimal. 

    \begin{figure}[htbp]
        \centering
        \includegraphics{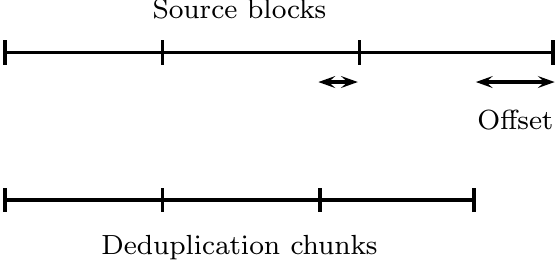}
        \caption{Loss of synchronization between source block and 
        deduplication chunk boundaries for fixed-length deduplication.}
        \label{fig:sync}
    \end{figure} 

    The reason for the bad performance of fixed-length deduplication is
    that the source blocks and the deduplication chunks are not properly
    synchronized.  Initially, the deduplication chunks are aligned with
    the source blocks.  However, whenever a source block of length $L+1$
    is observed, the deduplication chunks shift by one bit with respect
    to the source blocks (see Fig.~\ref{fig:sync}).  Over time, the
    boundary between deduplication chunks takes on all $L$ possible
    offsets with respect to the source block boundaries. Due to these
    $L$ possible starting points, the fixed-length deduplication scheme
    encounters $\Theta(L)$ distinct chunks instead of the only $A=2$
    distinct source symbols, resulting in the factor $\Theta(L) =
    \Theta(B)$ overhead compared to the optimal scheme. This argument is
    made precise in Appendix~\ref{sec:proofs_sync}.
\end{example}

From Example~\ref{eg:sync}, we see that the (general) deduplication
problem is fundamentally one of synchronizing deduplication chunks with
source blocks. Unfortunately, the fixed-length deduplication scheme
cannot achieve this synchronization. This motivates the use of the
variable-length deduplication scheme, described in
Section~\ref{sec:setting_schemes}, which utilizes anchor sequences to
achieve this synchronization. The next theorem bounds its performance.

\begin{theorem}
    \label{thm:variable}
    Consider the source model with $B$ source blocks drawn with
    replacement from the $A$ source symbols of expected length
    $\E(\msf{L})$. The performance of the variable-length deduplication
    scheme with optimized anchor length $M$ satisfies then
    \begin{equation*}
        1 \leq \frac{\Rv}{\Rs} \leq 
        1+ \frac{4B\bigl(1+\sqrt{\E(\msf{L})}\bigr) \log\bigl( B\E(\msf{L}) \bigr)}
        {\bigl(\min\{A,B\}(\E(\msf{L})-1) + (B-A)^+\log(A/2) - 2B\log(2\E(\msf{L}))\bigr)^+},
    \end{equation*}
    for $B$ large enough.
\end{theorem}
    
The proof of Theorem~\ref{thm:variable} is reported in
Appendix~\ref{sec:proofs_variable}. We illustrate this result with two examples.

\begin{example}[continues=eg:running]
    Consider again the scenario with $A = 10^5$ source symbols and with $B = 10^6$
    source blocks. This time, the source symbols are not of constant length, but
    have the same expected length $\E(\msf{L}) = 10^6$ bits as before.
    Theorem~\ref{thm:variable} shows then that variable-length deduplication
    has performance satisfying
    \begin{equation*}
       \Rs \leq \Rv \leq 2.6\Rs,
    \end{equation*}
    i.e., is within a factor $2.6$ of optimal. By numerically optimizing
    the value of the anchor length $M$ to minimize the upper
    bound~\eqref{eq:variable1} in Appendix~\ref{sec:proofs_variable}
    on the rate $\Rv$, this factor can be further reduced to $1.6$.
\end{example}

\begin{example}[continues=eg:sync]
    Consider again the scenario with $A=2$ source symbols with symbol-length
    distribution $P_\msf{L}$ assigning equal mass to the values $L$ and $L+1$
    with $L = B/3$. Recall that the fixed-length deduplication
    scheme had a rate at least order $B$ times larger than the optimal scheme:
    \begin{equation*}
        \frac{\Rf}{\Rs} \geq \Omega(B).
    \end{equation*}

    On the other hand, by tightening the arguments in the proof of
    Theorem~\ref{thm:variable} for the case $B > A^2$ (see
    Appendix~\ref{sec:proofs_sync2}), the rate of the variable-length
    deduplication scheme satisfies
    \begin{equation*}
        \frac{\Rv}{\Rs} \leq O(\log^3 B).
    \end{equation*}
    Thus, variable-length deduplication is only suboptimal by at most a
    polylogarithmic as opposed to a linear factor in this example. Thus,
    we see that variable-length deduplication is able to solve the problem
    of synchronizing source blocks and deduplication chunks, which was
    the cause of the poor performance of fixed-length deduplication.
\end{example}

\begin{figure}[htbp]
    \centering
    \includegraphics{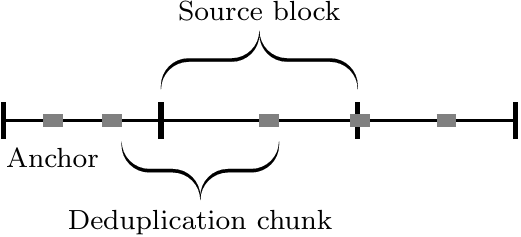}
    \caption{Source blocks and corresponding deduplication chunks for
    variable-length deduplication.}
    \label{fig:anchors}
\end{figure} 

The analysis in the proof of Theorem~\ref{thm:variable} in
Appendix~\ref{sec:proofs_variable} indicates that the optimal choice of the
anchor length $M$, governing the expected chunk length of the variable-length
deduplication scheme, balances two competing requirements.  On the one hand, for
each already encountered chunk, we need to encode a pointer into the dictionary.
A smaller chunk length increases both the number of chunks that need to be
encoded and the size of the pointers. On the other hand, consider chunks covering the
boundaries of two source blocks as shown in Fig.~\ref{fig:anchors}.
These boundary chunks will usually not be contained in the dictionary
(since there are $A^2$ such possible boundaries for $A$ source symbols), and
will have to be encoded directly. Hence, a larger chunk length increases the
amount of bits contained in inefficiently encodable boundary chunks. The choice
of anchor length $M \approx 0.5\log\E(\msf{L})$ in the proof of
Theorem~\ref{thm:variable} splits each source block into an average of about
$\sqrt{\E(\msf{L})}$ chunks of expected length about $\sqrt{\E(\msf{L})}$, which
balances these two detrimental effects. 

Unfortunately, even with this optimal choice of anchor length, using
deduplication chunks that have lengths of different order than the source blocks
can lead to asymptotically significantly suboptimal performance. This is
demonstrated with the next example.

\begin{example}
    \label{eg:duplicate}
    Consider the scenario with $A=\sqrt{B}$ source symbols of constant length
    $\msf{L} = L = \sqrt{B}$. From the preceding discussion, we see that this is
    the worst-case situation for variable-length deduplication, in which we can
    expect to see $\Theta(A^2)$ possible different boundary chunks. A slightly
    tightened version of Theorem~\ref{thm:variable} (which omits the last term
    in the denominator using that $\msf{L}$ is constant) together with a lower
    bound derived in Appendix~\ref{sec:proofs_duplicate}, show that then
    \begin{equation*}
        \Omega\bigl(B^{1/4}\log^{-2}(B)\bigr)
        \leq \frac{\Rv}{\Rs} 
        \leq O\bigl(B^{1/4}\bigr)
    \end{equation*}
    as $B\to\infty$. This statement has two implications: First, it
    shows that Theorem~\ref{thm:variable} is tight to within a
    polylogarithmic factor in $B$ for this setting; and second it shows
    that variable-length deduplication can still be polynomially
    suboptimal.
\end{example}

The multi-chunk deduplication scheme proposed in this paper circumvents
the competing requirements of how to choose the expected chunk length by
encoding multiple chunks jointly. This allows to choose the expected
chunk length to be quite small, thereby limiting the effect of the
boundary chunks, without the penalty of increased number of dictionary
pointers. The next theorem bounds the performance of this scheme.

\begin{theorem}
    \label{thm:multi}
    Consider the source model with $B$ source blocks drawn with replacement from
    the $A$ source symbols of expected length $\E(\msf{L})$. The performance of
    the multi-chunk deduplication scheme with optimized anchor length $M$
    satisfies then
    \begin{equation*}
        1 \leq \frac{\Rm}{\Rs} \leq
        1+O\biggl(\frac{B\log\bigl(AB\E(\msf{L})\bigr)}
        {\bigl(\min\{A,B\}(\E(\msf{L})-1) + (B-A)^+\log(A/2) - 2B\log(2\E(\msf{L}))\bigr)^+}\biggr)
    \end{equation*}
    as $B\to\infty$.
\end{theorem}

The proof of Theorem~\ref{thm:multi} is reported in
Appendix~\ref{sec:proofs_multi}. The theorem shows that, under the fairly mild
conditions $B^{\varepsilon} \leq A \leq (1-\varepsilon)B$ and $\E(\msf{L}) \leq
A^{\varepsilon/3}$ for some constant $\varepsilon > 0$, multi-chunk
deduplication is within a constant factor of optimal as $B\to\infty$.  Further,
if $A \leq B \leq o\bigl(A\E(\msf{L})/\log(A\E(\msf{L}))\bigr)$, then
multi-chunk deduplication is asymptotically optimal as $B\to\infty$.

\begin{example}[continues=eg:running]
    Consider again the scenario with $A = 10^5$ source symbols and with $B =
    10^6$ source blocks with expected length $\E(\msf{L}) = 10^6$ bits.
    Theorem~\ref{thm:multi} (using the explicit constant in the order notation
    from Appendix~\ref{sec:proofs_multi}) shows then that multi-chunk
    deduplication has performance satisfying
    \begin{equation*}
       \Rs \leq \Rm \leq 1.2\Rs.
    \end{equation*}
    By numerically optimizing the anchor length $M$
    to minimize the upper
    bound~\eqref{eq:mult_upper5} in Appendix~\ref{sec:proofs_multi}
    on the rate $\Rm$, this factor can be further
    reduced to $1.05$. Thus, the proposed multi-chunk deduplication scheme is
    quite close to optimal in this setting.
\end{example}

\begin{example}[continues=eg:duplicate]
    Consider again the scenario with $A=\sqrt{B}$ source symbols of constant
    length $\msf{L} = L = \sqrt{B}$. Recall that the variable-length
    deduplication scheme had a rate at least polynomially suboptimal:
    \begin{equation*}
        \frac{\Rv}{\Rs} 
        \geq \Omega(B^{1/4}\log^{-2}(B))
        \geq \Omega(B^{1/5}).
    \end{equation*}

    On the other hand, a slightly tightened version of Theorem~\ref{thm:multi},
    which omits the last term in the denominator using that $\msf{L}$ is
    constant, shows that the rate of the multi-chunk deduplication scheme
    satisfies
    \begin{equation*}
        \frac{\Rm}{\Rs} \leq O(1)
    \end{equation*}
    as $B\to\infty$. Thus, multi-chunk deduplication is order optimal in this
    case, as opposed to the polynomial loss factor of variable-length
    deduplication.
\end{example}

\section{Conclusion}
\label{sec:conclusion}

Motivated by the practical importance of data deduplication schemes but
a lack of theoretical results, this paper initiated the
information-theoretic analysis of these schemes. In order to enable this
analysis, it introduced a clean and tractable source model capturing the
long-range memory and large scale encountered in deduplication
applications. It then analyzed the two standard deduplication schemes
(fixed-length and variable-length). This analysis uncovers both the
strength and the weaknesses of both these schemes. The resulting insight
was used to construct a new scheme, called multi-chunk deduplication.
This new scheme was shown to be order optimal under fairly mild
assumptions.

While the description of the three deduplication schemes is general and
applies for any source sequence, their performance analysis
relies heavily on the specifics of the clean source model introduced in
this paper. In practice one does not have the luxury of such a clean
source model. However, the operation of the three deduplication schemes
in this paper depend only very weakly on the source model. Indeed, only
the chunk length $D$ or the anchor length $M$ need to be chosen. In
practice, one would tune these parameters empirically on a
representative dataset.

\appendices

\section{Analysis of Fixed-Length Deduplication with Constant Source-Symbol Length (Proof of Theorem~\ref{thm:fixed})}
\label{sec:proofs_fixed}

This appendix analyzes the rate of fixed-length deduplication for constant
source-symbol length. The length $\ell(\msf{S})$ of the source is in this case
the constant $BL$. Set the deduplication chunk length $D$ to be equal to the fixed
length $L$ of the source blocks. The deduplication chunks $\msf{Z}_1, \dots,
\msf{Z}_C$ are then equal to the source blocks $\msf{Y}_1, \dots, \msf{Y}_B$.

We start with an upper bound on the rate $\Rf$ of the fixed-length deduplication
scheme. The initial encoding of the length $\ell(\msf{S})=BL$ using a universal
code for the integers takes at most $2\log(BL)+3$ bits (see, e.g.,
\cite[Lemma~13.5.1]{cover06}). Consider then the encoding of some chunk $c$. The
flag indicating if the chunk is already in the dictionary takes one bit. If the
chunk is new, then it is added to the dictionary using $D$ bits. If the chunk is
already in the dictionary, then a pointer into the dictionary is encoded. Let
$\mc{Z}^{c-1}$ be the dictionary when processing chunk $c$. Then this encoded pointer takes at
most $\log\card{\mc{Z}^{c-1}}+1$ bits. 

The expected rate of fixed-length deduplication is thus upper bounded by
\begin{align*}
    \Rf
    & \leq 2\log(BL)+3
    + \sum_{c=1}^C \E\Bigl( 
    1 + \ind_{\{\msf{Z}_c\notin\mc{Z}^{c-1}\}} D
    + \ind_{\{\msf{Z}_c\in\mc{Z}^{c-1}\}} (\log\card{\mc{Z}^{c-1}}+1) 
    \Bigr) \\
    & \leq 2\log(BL)+3+2C
    + \sum_{c=1}^C\Bigl( 
    D \Pp(\msf{Z}_c\notin\mc{Z}^{c-1})
    + A^{-1}\E\bigl( \card{\mc{Z}^{c-1}}\log\card{\mc{Z}^{c-1}} \bigr)
    \Bigr),
\end{align*}
where $\ind_{\{\cdot\}}$ denotes the indicator function, and where we have used that
\begin{align*}
    \E\bigl( \ind_{\{\msf{Z}_c\in\mc{Z}^{c-1}\}} \log\card{\mc{Z}^{c-1}} \bigr)
    & = \E\Bigl( \E\bigl( \ind_{\{\msf{Z}_c\in\mc{Z}^{c-1}\}} \log\card{\mc{Z}^{c-1}} \bigm\vert \card{\mc{Z}^{c-1}} \bigr) \Bigr) \\
    & = \E\Bigl( \Pp\bigl( \msf{Z}_c\in\mc{Z}^{c-1} \bigm\vert \card{\mc{Z}^{c-1}} \bigr) \log\card{\mc{Z}^{c-1}}  \Bigr) \\
    & = A^{-1}\E\bigl( \card{\mc{Z}^{c-1}} \log\card{\mc{Z}^{c-1}} \bigr).
\end{align*}
Using that $C=B$, $D=L$, and $\msf{Z}_c = \msf{Y}_c$, this upper bound can be
rewritten as
\begin{equation}
    \label{eq:fixed_upper1}
    \Rf
    \leq 2\log(BL)+3+2B
    + \sum_{b=1}^B\Bigl( 
    L \Pp(\msf{Y}_b\notin\mc{Y}^{b-1})
    + A^{-1}\E\bigl( \card{\mc{Y}^{b-1}}\log\card{\mc{Y}^{b-1}} \bigr)
    \Bigr),
\end{equation}
where 
\begin{equation}
    \label{eq:mcydef}
    \mc{Y}^b \defeq \{\msf{Y}_1, \msf{Y}_2, \dots, \msf{Y}_b\}
\end{equation}
denotes the set of all \emph{distinct} source blocks seen up to block $b$. 

We continue with a lower bound on the rate $\Rs$ of the optimal code. Since the
code is prefix free, its rate is lower bounded as
\begin{equation}
    \label{eq:fixed_lower1}
    \Rs \geq H(\msf{S})
\end{equation}
(see, e.g., \cite[Theorem~5.4.1]{cover06}). As the source blocks are of constant
length, we have
\begin{equation}
    \label{eq:fixed_lower2}
    H(\msf{S}) 
    = H(\msf{Y}^B)
    = \sum_{b=1}^B H(\msf{Y}_b \mid \msf{Y}^{b-1})
\end{equation}
with 
\begin{equation*}
    \msf{Y}^{b-1} \defeq (\msf{Y}_1, \msf{Y}_2, \dots, \msf{Y}_{b-1}).
\end{equation*}
Each term in the sum on the right-hand side satisfies
\begin{align}
    \label{eq:fixed_lower3}
    H(\msf{Y}_b \mid \msf{Y}^{b-1})
    & \geq H\bigl(\msf{Y}_b \bigm\vert \msf{Y}^{b-1}, \ind_{\{\msf{Y}_b\in\mc{Y}^{b-1}\}} \bigr) \notag\\
    & = \Pp(\msf{Y}_b\notin\mc{Y}^{b-1}) H\bigl(\msf{Y}_b \bigm\vert \msf{Y}^{b-1}, \msf{Y}_b\notin\mc{Y}^{b-1} \bigr) 
    + \Pp(\msf{Y}_b\in\mc{Y}^{b-1}) H\bigl(\msf{Y}_b \bigm\vert \msf{Y}^{b-1}, \msf{Y}_b\in\mc{Y}^{b-1} \bigr).
\end{align}
Conditioned on $\msf{Y}_b\notin\mc{Y}^{b-1}$ and $\msf{Y}^{b-1}$, the
source block $\msf{Y}_b$ is uniformly distributed over
$\{0,1\}^L\setminus\mc{Y}^{b-1}$. Hence,
\begin{align*}
    H\bigl(\msf{Y}_b \bigm\vert \msf{Y}^{b-1}, \msf{Y}_b\notin\mc{Y}^{b-1} \bigr)
    & = \E\bigl( \log\card{\{0,1\}^L\setminus\mc{Y}^{b-1}} \bigm\vert \msf{Y}_b\notin\mc{Y}^{b-1} \bigr)
    \\
    & = \E\bigl( \log(2^L-\card{\mc{Y}^{b-1}}) \bigm\vert \msf{Y}_b\notin\mc{Y}^{b-1} \bigr) \\
    & \overset{(a)}{\geq} \log(2^L-A) \\
    & \overset{(b)}{\geq} L-1
\end{align*}
using that $\card{\mc{Y}^{b-1}} \leq A$ for $(a)$ and that $A \leq 2^{L-1}$ by
assumption for $(b)$. This implies that
\begin{equation}
    \label{eq:fixed_lower4}
    \Pp(\msf{Y}_b\notin\mc{Y}^{b-1})H\bigl(\msf{Y}_b \bigm\vert \msf{Y}^{b-1}, \msf{Y}_b\notin\mc{Y}^{b-1} \bigr)
    \geq \Pp(\msf{Y}_b\notin\mc{Y}^{b-1})(L-1).
\end{equation}
Conditioned on $\msf{Y}_b\in\mc{Y}^{b-1}$ and $\msf{Y}^{b-1}$, the source block
$\msf{Y}_b$ is uniformly distributed over $\mc{Y}^{b-1}$. Hence,
\begin{align}
    \label{eq:fixed_lower5}
    \Pp(\msf{Y}_b\in\mc{Y}^{b-1}) & H\bigl(\msf{Y}_b \bigm\vert \msf{Y}^{b-1}, \msf{Y}_b\in\mc{Y}^{b-1} \bigr) \notag\\
    & = \Pp(\msf{Y}_b\in\mc{Y}^{b-1}) H\bigl(\msf{Y}_b \bigm\vert \msf{Y}^{b-1}, \card{\mc{Y}^{b-1}}, \msf{Y}_b\in\mc{Y}^{b-1} \bigr) \notag\\
    & = \sum_{a=1}^A \Pp(\card{\mc{Y}^{b-1}} = a) \Pp(\msf{Y}_b\in\mc{Y}^{b-1} \mid \card{\mc{Y}^{b-1}} = a) H\bigl(\msf{Y}_b \bigm\vert \msf{Y}^{b-1}, \card{\mc{Y}^{b-1}} = a, \msf{Y}_b\in\mc{Y}^{b-1} \bigr) \notag\\
    & = \sum_{a=1}^A \Pp(\card{\mc{Y}^{b-1}} = a)aA^{-1} \log(a) \notag\\
    & = A^{-1}\E\bigl( \card{\mc{Y}^{b-1}} \log\card{\mc{Y}^{b-1}} \bigr).
\end{align}
Combining~\eqref{eq:fixed_lower1}--\eqref{eq:fixed_lower5} yields
\begin{equation}
    \label{eq:fixed_lower6}
    \Rs 
    \geq \sum_{b=1}^B \Bigl(
    (L-1) \Pp(\msf{Y}_b\notin\mc{Y}^{b-1})
    + A^{-1}\E\bigl( \card{\mc{Y}^{b-1}}\log\card{\mc{Y}^{b-1}} \bigr)
    \Bigr).
\end{equation}

To obtain a more explicit expression, we further lower bound $\Rs$ as
\begin{align}
    \label{eq:fixed_lower7}
    \Rs 
    & \overset{(a)}{\geq} \sum_{b=1}^B \Bigl(
    (L-1) (1-(b-1)/A)^+
    + A^{-1}\E\card{\mc{Y}^{b-1}}\log\E\card{\mc{Y}^{b-1}}
    \Bigr) \notag\\
    & \overset{(b)}{=} (L-1)\sum_{b=1}^B (1-(b-1)/A)^+
    + \sum_{b=1}^B\bigl(1-(1-1/A)^{b-1}\bigr)\log\bigl( A\bigl(1-(1-1/A)^{b-1}\bigr) \bigr) \notag\\
    & \overset{(c)}{\geq} 0.5(L-1)\min\{A,B\} + 0.5(B-A)^+\log(A/2),
\end{align}
where $(a)$ follows from $\card{\mc{Y}^{b-1}}\leq b-1$ and from  the convexity
of $x\log x$ and Jensen's inequality, $(b)$ follows from
\begin{equation*}
    \E\card{\mc{Y}^{b-1}} = A\bigl(1-(1-1/A)^{b-1}\bigr)
\end{equation*}
since the $\msf{Y}^{b-1}$ are chosen uniformly with replacement from the set
$\mc{X}$ of cardinality $A$, and $(c)$ follows from
\begin{align*}
    \sum_{b=1}^B (1-(b-1)/A)^+ 
    & = 
    \begin{cases}
        B(2A-B+1)/(2A), & \text{if $B\leq A$}, \\
        (A+1)/2, & \text{if $B > A$}.
    \end{cases} \\
    & \geq 0.5\min\{A, B\}
\end{align*}
and from
\begin{equation*}
    1-(1-1/A)^{b-1} \geq 1-\exp(-1) \geq 0.5
\end{equation*}
for $b \geq A+1$ and
\begin{equation*}
    \bigl(1-(1-1/A)^{b-1}\bigr)\log\bigl( A\bigl(1-(1-1/A)^{b-1}\bigr) \bigr)
    \geq 0
\end{equation*}
for $1 \leq b \leq A$.

From~\eqref{eq:fixed_upper1} and~\eqref{eq:fixed_lower6}, we obtain
\begin{align*}
    \Rf-\Rs 
    & \leq 2\log(BL)+3+2B+\sum_{b=1}^B \Pp(\msf{Y}_b\notin\mc{Y}^{b-1}) \\
    & \leq 2\log(BL)+3+3B.
\end{align*}
Combining this with~\eqref{eq:fixed_lower7} yields
\begin{equation*}
    \frac{\Rf}{\Rs}-1
    \leq \frac{2\log(BL)+3+3B}{0.5(L-1)\min\{A,B\} + 0.5(B-A)^+\log(A/2)}.
\end{equation*}
For $B$ large enough, this can be simplified as
\begin{equation*}
    \frac{\Rf}{\Rs}-1
    \leq 7\frac{\log(L)+B}{(L-1)\min\{A,B\} + (B-A)^+\log(A/2)},
\end{equation*}
concluding the proof. \hfill\IEEEQED

\section{Analysis of Fixed-Length Deduplication with Variable Source-Symbol Length (Example~\ref{eg:sync})}
\label{sec:proofs_sync}

This appendix contains the formal analysis for Example~\ref{eg:sync}.
We start with an upper bound on $\Rs$. Since $\Rs$ is the rate of the optimal
prefix-free code, it is upper bounded as
\begin{equation*}
    \Rs \leq H(\msf{S})+1
\end{equation*}
(see, e.g., \cite[Theorem~5.4.1]{cover06}). Now,
\begin{align*}
    H(\msf{S})
    & \leq H\bigl( \msf{S}, \msf{Y}^B \bigr) \notag\\
    & = H(\msf{Y}^B)+H( \msf{S} \mid \msf{Y}^B \bigr) \notag\\
    & = H(\msf{Y}^B) \notag\\
    & \leq H(\msf{Y}^B, \mc{X}) \notag\\
    & = H(\mc{X})+H(\msf{Y}^B \mid \mc{X}) \notag\\
    & \leq \sum_{a=1}^A \bigl( H(\msf{L}_a) + H(\msf{X}_a \mid \msf{L}_a) \bigr)
    +\sum_{b=1}^B H(\msf{Y}_b \mid \mc{X}) \notag\\
    & \leq A(L+2)+B\log(A).
\end{align*}
Thus, 
\begin{equation}
    \label{eq:fixed2_upper1}
    \Rs \leq A(L+2)+B\log(A)+1.
\end{equation}

\begin{figure}[htbp]
    \centering
    \subfigure[Chunk offsets\label{fig:offset_a}]{\includegraphics{offset.pdf}} \\
    \subfigure[Markov chain for $L=4$\label{fig:offset_b}]{\includegraphics{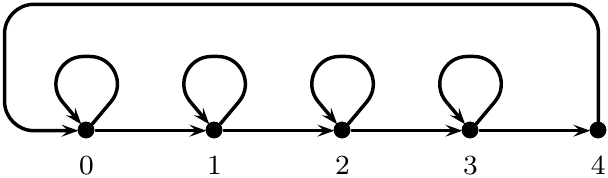}}
    \caption{Evolution of the offset between source block and deduplication
    chunk boundaries (a) and corresponding Markov chain (b).}
    \label{fig:offset}
\end{figure} 

We continue with a lower bound on the rate of fixed-length deduplication. We set
the chunk length $D$ to be equal to $L$. Since the source blocks have lengths
either $L$ or $L+1$, the deduplication chunk boundaries may no longer be aligned
with the source block boundaries. Define the offset of the current
chunk as the distance from the end of that chunk to the start of the next source block
(see Fig.~\ref{fig:offset_a}).

Recall that $A=2$, and assume for the moment that the source alphabet $\mc{X}$
has one source symbol of length $L$ and one of length $L+1$ (this happens with
probability $1/2$).  The evolution of this offset is then governed by a Markov
chain with $L+1$ states as depicted in Fig.~\ref{fig:offset_b}. The initial
state is $0$ and all outgoing edges of a state have uniform probability. Observe
that in each state we make a transition to the right (modulo $L+1$) with
probability at least $1/2$ or stay in the current state with probability at most
$1/2$. Hence, after $2L$ transitions, we have traversed the chain at least once
in expectation. 

Consider now the two source symbols $\msf{X}_1$ and $\msf{X}_2$. And assume for
the moment that $\ell(\msf{X}_1) = L$. By the law of large numbers, about $1/4$
of the deduplication chunks will be substrings of $\msf{X}_1\msf{X}_1$ (the
concatenation of $\msf{X}_1$ with itself) with high probability for $L$ large
enough. Consider all possible chunks of length $L$ starting with different
offsets in $\msf{X}_1\msf{X}_1$. We next argue that with high probability all
these $L$ different chunks are unique.

By~\cite[Example~10.5]{lint01}, there are
\begin{equation*}
    \sum_{d \mid L}\mu(d)2^{L/d}
\end{equation*}
binary sequences of length $L$ for which all circular shifts are distinct (these
are called aperiodic necklaces in Combinatorics), where $\mu(d)\in\{0,\pm 1\}$
is the M\"obius function (see, e.g.,~\cite[p.~92]{lint01} for a definition).
Since $\mu(1) = 1$ and $\mu(d) \geq -1$, we can lower bound this as
\begin{align*}
    \sum_{d \mid L}\mu(d)2^{L/d}
    & \geq  2^L- \sum_{d \mid L, d > 1}2^{L/d} \\
    & \geq  2^L- 2^{L/2+1} \\
    & = 2^L(1-o(1)).
\end{align*}
This shows that, as $L\to\infty$, the vast majority of binary sequences have the
property that all their circular shifts are distinct. In particular, with high
probability $\msf{X}_1$ will have this property.

Putting these arguments together, we obtain the following.  With probability
$1/2$ the two source symbols have distinct lengths. With probability at least
$1/2$, the shorter of the two source symbols (the one with length $L$) will have
distinct circular shifts for $L$ large enough. If $B = 3L$ and $L$ large enough,
then we will see every possible deduplication chunk offset at least once with
probability at least $1/2$.  Moreover, with probability $1/2$ at least $L/8$ of
the deduplication chunks will contain circular shifts of the shorter source
symbol. Since each of these are distinct, they will all have to be entered into
the dictionary, using $L$ bits each. Thus,
\begin{equation}
    \label{eq:fixed2_lower1}
    \Rf \geq 2^{-7}L^2.
\end{equation}

Combining~\eqref{eq:fixed2_upper1} (with $A=2$ and $B=3L$) and
\eqref{eq:fixed2_lower1} shows that
\begin{equation*}
    \frac{\Rf}{\Rs} \geq \frac{2^{-7}L^2}{5L+5} = \Omega(L) = \Omega(B)
\end{equation*}
as claimed. \hfill\IEEEQED

\section{Analysis of Variable-Length Deduplication (Proof of Theorem~\ref{thm:variable})}
\label{sec:proofs_variable}

We start with an upper bound on the rate $\Rv$ of variable-length deduplication.
The initial encoding of the length $\ell(\msf{S})$ using a universal code for
the integers takes at most $2\log\ell(\msf{S})+3$ bits (see again
\cite[Lemma~13.5.1]{cover06}). Consider then the encoding of chunk $c$. The flag
indicating if the chunk is already in the dictionary takes one bit. If the chunk
is new, then it is added to the dictionary using $\ell(\msf{Z}_c)$ bits. If the
chunk is already in the dictionary, then a pointer into the dictionary is
encoded. Let again $\mc{Z}^{c-1}$ be the dictionary when encoding chunk $c$.
Then this encoded pointer takes at most $\log\card{\mc{Z}^{c-1}}+1$ bits.

Let $\Rv(\msf{S})$ be the rate of variable-length deduplication for a particular
source sequence $\msf{S}$, so that $\Rv = \E\bigl(\Rv(\msf{S})\bigr)$.
The rate $\Rv(\msf{S})$ is then upper bounded by
\begin{align}
    \label{eq:variable_upper1}
    \Rv(\msf{S})
    & \leq 2\log\ell(\msf{S})+3
    + \sum_{c=1}^\msf{C}\Bigl( 
    1 + \ind_{\{\msf{Z}_c\notin\mc{Z}^{c-1}\}} \ell(\msf{Z}_c) 
    + \ind_{\{\msf{Z}_c\in\mc{Z}^{c-1}\}} \bigl(\log\card{\mc{Z}^{c-1}}+1\bigr) \Bigr) \notag\\
    & \leq 2\log\ell(\msf{S})+3+2\msf{C}
    + \sum_{c=1}^\msf{C}\ind_{\{\msf{Z}_c\notin\mc{Z}^{c-1}\}} \ell(\msf{Z}_c) 
    + \sum_{c=1}^\msf{C}\ind_{\{\msf{Z}_c\in\mc{Z}^{c-1}\}} \log\card{\mc{Z}^{c-1}}.
\end{align}

Now, we have two distinct parsings of the source sequence $\msf{S}$. The first
is defined by the source blocks, $\msf{Y}_1, \msf{Y}_2, \dots, \msf{Y}_B$; the
second is defined by the deduplication chunks $\msf{Z}_1, \msf{Z}_2, \dots,
\msf{Z}_{\msf{C}}$. We would like to relate these two parsings. To this end, let
$\mc{C}_b$ denote the indices of those chunks from $\msf{S}$ starting in
$\msf{Y}_b$. We say that a chunk $\msf{Z}_c$ is in $\msf{Y}_b$ if
$c\in\mc{C}_b$. 

\begin{example}
    \label{eg:anchors}
    Consider $\msf{Y}_1 = 100110110001$, $\msf{Y}_2 = 01110010$,
    $\msf{Y}_3 = 010011$ so that the source sequence is $\msf{S} =
    10011011000101110010010011$.  The parsing of $\msf{S}$ into chunks
    with anchor $0^M = 00$ yields $\msf{Z}_1 = 100$, $\msf{Z}_2 =
    1101100$, $\msf{Z}_3 = 01011100$, $\msf{Z}_4 = 100$, $\msf{Z}_5 =
    100$, $\msf{Z}_6 = 11$. This situation is depicted in
    Fig.~\ref{fig:anchors} in Section~\ref{sec:results}.  In this
    setting $\mc{C}_1 = \{1, 2, 3\}$, $\mc{C}_2 = \{4\}$, and $\mc{C}_3
    = \{5, 6\}$. 
\end{example}

Consider the chunks in source block $\msf{Y}_b$. As
Fig.~\ref{fig:anchors} in Section~\ref{sec:results} shows, some of them
depend on the values of the neighboring source blocks $\msf{Y}_{b-1}$
and $\msf{Y}_{b+1}$. We call these the ``boundary'' chunks of
$\msf{Y}_b$ and denote their indices by $\partial\mc{C}_b$. Other chunks
in $\msf{Y}_b$ are the same irrespective of the values of the
neighboring source blocks. We call these the ``interior'' chunks of
$\msf{Y}_b$ and denote their indices by $\mc{C}_b^\circ$. Formally, for
a source block $\msf{Y}_b$, we say that $\msf{Z}_c$ with $c\in\mc{C}_b$
is an interior chunk if it appears in the variable-length chunking of
every sequence $y_{b-1}\msf{Y}_b y_{b+1}$ with $y_{b-1},
y_{b+1}\in\{0,1\}^*$. Any chunk $\msf{Z}_c$ that is not an interior
chunk is defined to be a boundary chunk.  By definition, the interior
chunks thus have the property that if $\msf{Y}_b = \msf{Y}_{\tilde{b}}$,
then $\{\msf{Z}_c: c\in\mc{C}_b^\circ\} = \{\msf{Z}_c:
c\in\mc{C}_{\tilde{b}}^\circ\}$. Note that this last conclusion does
generally not hold for $\partial\mc{C}_b$ and
$\partial\mc{C}_{\tilde{b}}$.

\begin{example}[continues=eg:anchors]
    In this setting we have
    $\mc{C}_1^\circ = \{2\}$, $\mc{C}_2^\circ = \emptyset$, $\mc{C}_3^\circ =
    \emptyset$ and $\partial\mc{C}_1 = \{1,3\}$, $\partial\mc{C}_2 = \{4\}$,
    $\partial\mc{C}_3 = \{5, 6\}$.
\end{example}

\begin{figure}[htbp]
    \centering
    \subfigure[\label{fig:boundary0}]{\includegraphics{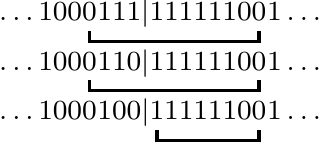}} \\
    \subfigure[\label{fig:boundary1}]{\includegraphics{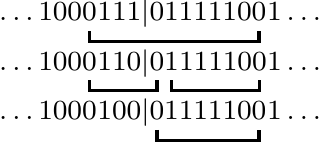}} \\
    \subfigure[\label{fig:boundary2}]{\includegraphics{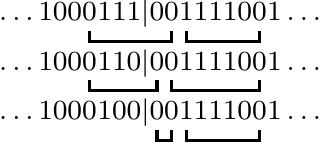}} \\
    \subfigure[\label{fig:boundary3}]{\includegraphics{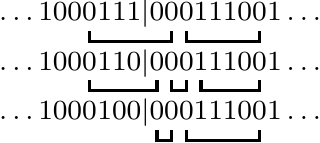}} \\
    \subfigure[\label{fig:boundary4}]{\includegraphics{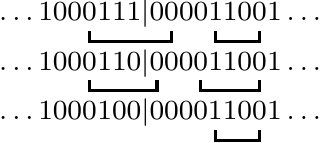}}
    \caption{Boundary chunks (indicated by underbraces) as defined by
        variable-length deduplication for anchor length $M=2$.
        The vertical line $\vert$ indicates the boundary between the source
        blocks $\msf{Y}_{b-1}$ and $\msf{Y}_b$.}
    \label{fig:boundary}
\end{figure} 

Usually, $\partial\mc{C}_b$ contains only one chunk index, which corresponds to
the final chunk starting in $\msf{Y}_b$ but ending in $\msf{Y}_{b+1}$ (see
Figs.~\ref{fig:boundary0} and~\ref{fig:boundary1}). However, $\partial\mc{C}_b$
can contain additional chunk indices. In particular, if the boundary between
$\msf{Y}_{b-1}$ and $\msf{Y}_{b}$ forms an anchor sequence, then the first chunk
starting in $\msf{Y}_b$ may also be a boundary chunk (see
Figs.~\ref{fig:boundary1}--\ref{fig:boundary4}). Finally, if $\msf{Y}_b$ starts
with between $M+1$ and $2M-1$ zeros (where $M$ is the anchor length), then it
may contain a third boundary chunk consisting of the anchor sequence $0^M$ by
itself (see Fig.~\ref{fig:boundary3}). Observe that when $\msf{Y}_b$ starts with
$2M$ or more zeros, then there is always a $0^M$ chunk, irrespective of the
value of $\msf{Y}_{b-1}$, and therefore $0^M$ is not a boundary chunk in this
case (see Fig.~\ref{fig:boundary4}). In general, we thus have
$\card{\partial\mc{C}_b}\leq 3$. We will later choose $M$ such that
$\E\card{\mc{C}_b} = \omega(1)$ as $\E(\msf{L})\to\infty$, in which case the vast
majority of indices in $\mc{C}_b$ will correspond to interior chunks.

Let us return to the upper bound~\eqref{eq:variable_upper1} for $\Rv(\msf{S})$,
and consider the first sum corresponding to new chunks. We can now rewrite this
sum as
\begin{align}
    \label{eq:variable_upper2}
    \sum_{c=1}^\msf{C}\ind_{\{\msf{Z}_c\notin\mc{Z}^{c-1}\}} \ell(\msf{Z}_c) 
    & = \sum_{b=1}^B\sum_{c\in\mc{C}_b}\ind_{\{\msf{Z}_c\notin\mc{Z}^{c-1}\}} \ell(\msf{Z}_c) \notag\\
    & = \sum_{b=1}^B\sum_{c\in\mc{C}_b^\circ}\ind_{\{\msf{Z}_c\notin\mc{Z}^{c-1}\}} \ell(\msf{Z}_c) 
    + \sum_{b=1}^B\sum_{c\in\partial\mc{C}_b}\ind_{\{\msf{Z}_c\notin\mc{Z}^{c-1}\}} \ell(\msf{Z}_c).
\end{align}
As before, denote by $\mc{Y}^b$ the set of all distinct source blocks seen up to
block $b$ as defined in~\eqref{eq:mcydef} in Appendix~\ref{sec:proofs_fixed}.
Note that if $\msf{Z}_c\notin\mc{Z}^{c-1}$ for \emph{any} $c\in\mc{C}_b^\circ$,
then $\msf{Y}_b\notin\mc{Y}^{b-1}$. Hence
\begin{align*}
    \sum_{b=1}^B\sum_{c\in\mc{C}_b^\circ}\ind_{\{\msf{Z}_c\notin\mc{Z}^{c-1}\}} \ell(\msf{Z}_c) 
    & \leq \sum_{b=1}^B\ind_{\{\msf{Y}_b\notin\mc{Y}^{b-1}\}}\sum_{c\in\mc{C}_b^\circ} \ell(\msf{Z}_c) \\
    & \leq \sum_{b=1}^B\ind_{\{\msf{Y}_b\notin\mc{Y}^{b-1}\}} \ell(\msf{Y}_b),
\end{align*}
where we have used that
\begin{equation*}
    \sum_{c\in\mc{C}_b^\circ}\ell(\msf{Z}_c) \leq \ell(\msf{Y}_b).
\end{equation*}
Substituting this into~\eqref{eq:variable_upper2} yields
\begin{equation}
    \label{eq:variable_upper3}
    \sum_{c=1}^\msf{C}\ind_{\{\msf{Z}_c\notin\mc{Z}^{c-1}\}} \ell(\msf{Z}_c) 
    \leq \sum_{b=1}^B\ind_{\{\msf{Y}_b\notin\mc{Y}^{b-1}\}} \ell(\msf{Y}_b)
    + \sum_{b=1}^B\sum_{c\in\partial\mc{C}_b}\ind_{\{\msf{Z}_c\notin\mc{Z}^{c-1}\}}\ell(\msf{Z}_c).
\end{equation}

Consider then the second sum in~\eqref{eq:variable_upper1} corresponding to old
chunks. We can upper bound this sum as
\begin{equation}
    \label{eq:variable_upper5}
    \sum_{c=1}^\msf{C}\ind_{\{\msf{Z}_c\in\mc{Z}^{c-1}\}} \log\card{\mc{Z}^{c-1}}
    \leq \msf{C}\log\card{\mc{Z}^{\msf{C}}}.
\end{equation}

Substituting~\eqref{eq:variable_upper3} and~\eqref{eq:variable_upper5} into
\eqref{eq:variable_upper1} yields
\begin{align*}
    \Rv(\msf{S})
    & \leq 2\log\ell(\msf{S})+3+2\msf{C}
    + \sum_{b=1}^B\ind_{\{\msf{Y}_b\notin\mc{Y}^{b-1}\}}\ell(\msf{Y}_b)
    + \sum_{b=1}^B\sum_{c\in\partial\mc{C}_b}\ind_{\{\msf{Z}_c\notin\mc{Z}^{c-1}\}}\ell(\msf{Z}_c)
    + \msf{C}\log\card{\mc{Z}^{\msf{C}}}.
\end{align*}
Taking expectations on both sides results in an upper bound on $\Rv$:
\begin{align}
    \label{eq:variable_upper6}
    \Rv
    & \leq 2\E\log\ell(\msf{S})+3+2\E(\msf{C})
    + \sum_{b=1}^B\E\bigl(\ind_{\{\msf{Y}_b\notin\mc{Y}^{b-1}\}}\ell(\msf{Y}_b)\bigr) \notag\\
    & \quad {} + \sum_{b=1}^B\E\biggl(\sum_{c\in\partial\mc{C}_b}\ind_{\{\msf{Z}_c\notin\mc{Z}^{c-1}\}}\ell(\msf{Z}_c)\biggr) 
    + \E\bigl(\msf{C}\log\card{\mc{Z}^{\msf{C}}}\bigr).
\end{align}
We now upper bound each of these expectations in turn.

The first expectation in~\eqref{eq:variable_upper6} is upper bounded as
\begin{equation}
    \label{eq:variable_upper7}
    \E\log\ell(\msf{S})
    \leq \log\E\ell(\msf{S}) 
    = \log\bigl(B\E(\msf{L})\bigr)
\end{equation}
using Jensen's inequality.

For the second expectation in~\eqref{eq:variable_upper6}, observe that
the number of chunks starting in source block $\msf{Y}_b$ is at most $1$
plus the number of times the anchor $0^M$ appears in $\msf{Y}_b$ alone
(see again Fig.~\ref{fig:anchors} in Section~\ref{sec:results}). Since
the expectation of that latter number is upper bounded by
$2^{-M}\E(\msf{L})$, we obtain
\begin{equation}
    \label{eq:variable_upper8}
    \E(\msf{C}) 
    = \sum_{b=1}^B\E\card{\mc{C}_b}
    \leq B\bigl(1+2^{-M}\E(\msf{L})\bigr).
\end{equation}

The third expectation in~\eqref{eq:variable_upper6} is equal to
\begin{equation}
    \label{eq:variable_upper9}
    \sum_{b=1}^B\E\bigl(\ind_{\{\msf{Y}_b\notin\mc{Y}^{b-1}\}}\ell(\msf{Y}_b)\bigr)
    = \E(\msf{L})\sum_{b=1}^B \Pp(\msf{Y}_b\notin\mc{Y}^{b-1}),
\end{equation}
where we have used the independence of $\ind_{\{\msf{Y}_b\notin\mc{Y}^{b-1}\}}$ and
$\ell(\msf{Y}_b)$. 

Consider next the fourth expectation
\begin{equation*}
    \sum_{b=1}^B\E\biggl(\sum_{c\in\partial\mc{C}_b}\ind_{\{\msf{Z}_c\notin\mc{Z}^{c-1}\}}\ell(\msf{Z}_c)\biggr)
    \leq \sum_{b=1}^B\E\biggl(\sum_{c\in\partial\mc{C}_b}\ell(\msf{Z}_c)\biggr)
\end{equation*}
in~\eqref{eq:variable_upper6}. Consider the boundary chunks arising at the
boundary between $\msf{Y}_b$ and $\msf{Y}_{b+1}$ (see again
Fig.~\ref{fig:boundary}). The total number of bits due to those chunks is upper
bounded by the sum of two terms: The number of bits from the end of $\msf{Y}_b$
backwards until the end of the first (counting backwards) occurrence of $0^M$
(or $\ell(\msf{Y}_b)$ if no such match exists in $\msf{Y}_b$). Plus the number
of bits from the start of $\msf{Y}_{b+1}$ forwards until the end of the first
(counting forwards) occurrence of $10^M$ (or $\ell(\msf{Y}_{b+1})$ if no such
match exists in $\msf{Y}_{b+1}$).

Note that a source symbol $\msf{X}_a$ is (by itself) a $\Bernoulli(1/2)$
process of random length. The expected value of the end of the first occurrence
of $10^M$ in an infinite-length $\Bernoulli(1/2)$ process is $2^{M+1}$ by
\cite[Theorem~8.3]{sedgewick13}. The expected value of the end of the first
occurrence of $0^M$ in an infinite-length $\Bernoulli(1/2)$ process is
$2^{M+1}-2$ by \cite[Theorem~8.2]{sedgewick13}. Hence
\begin{equation}
    \label{eq:variable_upper11}
    \sum_{b=1}^B\E\biggl(\sum_{c\in\partial\mc{C}_b}\ind_{\{\msf{Z}_c\notin\mc{Z}^{c-1}\}}\ell(\msf{Z}_c)\biggr) 
    \leq B2^{M+2}.
\end{equation}
In this bound, the event that some source blocks may not contain an anchor
sequence is captured by the event that the match of the anchor sequence in the
infinite-length $\Bernoulli(1/2)$ process is beyond the length
$\ell(\msf{Y}_b)$ of the corresponding source block. 

Consider then the last expectation 
\begin{align*}
    \E\bigl(\msf{C}\log\card{\mc{Z}^{\msf{C}}}\bigr)
\end{align*}
in~\eqref{eq:variable_upper6}. We have
\begin{align*}
    \card{\mc{Z}^{\msf{C}}} \leq \ell(\msf{S}) \leq 2B\E(\msf{L}).
\end{align*}
Using this, we can upper bound 
\begin{align}
    \label{eq:variable_upper10}
    \E\bigl(\msf{C}\log\card{\mc{Z}^{\msf{C}}}\bigr)
    & \leq B\bigl(1+2^{-M}\E(\msf{L})\bigr) \log\bigl(2B\E(\msf{L})\bigr),
\end{align}
where we have used~\eqref{eq:variable_upper8}.

Substituting~\eqref{eq:variable_upper7}--\eqref{eq:variable_upper10} into
\eqref{eq:variable_upper6} results in
\begin{align}
    \label{eq:variable_upper13}
    \Rv
    & \leq 
    2\log\bigl(B\E(\msf{L})\bigr)
    +3
    +2B\bigl(1+2^{-M}\E(\msf{L})\bigr)
    +\E(\msf{L})\sum_{b=1}^B \Pp(\msf{Y}_b\notin\mc{Y}^{b-1}) \notag\\
    & \quad {} + B2^{M+2}
    + B\bigl(1+2^{-M}\E(\msf{L})\bigr) \log\bigl( 2B\E(\msf{L}) \bigr).
\end{align}

We next derive a lower bound on $\Rs$. As before, we have
\begin{equation}
    \label{eq:variable_lower1}
    \Rs \geq H(\msf{S}).
\end{equation}
Now, 
\begin{align}
    \label{eq:variable_lower2}
    H(\msf{S})
    & = H\bigl( \msf{S}, \ell(\msf{Y}_1), \dots, \ell(\msf{Y}_B) \bigr)
    - H\bigl( \ell(\msf{Y}_1), \dots, \ell(\msf{Y}_B) \bigm\vert \msf{S} \bigr) \notag\\
    & \geq H(\msf{Y}^B)- H\bigl( \ell(\msf{Y}_1), \dots, \ell(\msf{Y}_B) \bigr) \notag\\
    & \geq H(\msf{Y}^B)- BH(\msf{L}).
\end{align}
The term $BH(\msf{L})$ can be bounded as
\begin{equation}
    \label{eq:variable_lower3a}
    BH(\msf{L})
    \leq B(1+\log\E(\msf{L}))
\end{equation}
using $1 \leq \msf{L} \leq 2\E(\msf{L})$.  The term $H(\msf{Y}^B)$ can be bounded
similarly to \eqref{eq:fixed_lower3} in Appendix~\ref{sec:proofs_fixed} as
\begin{equation}
    \label{eq:variable_lower3}
    H(\msf{Y}^B) = \sum_{b=1}^B H(\msf{Y}_b \mid \msf{Y}^{b-1})
\end{equation}
with
\begin{align}
    \label{eq:variable_lower4}
    H(\msf{Y}_b \mid \msf{Y}^{b-1})
    & \geq \Pp(\msf{Y}_b\notin\mc{Y}^{b-1}) H\bigl(\msf{Y}_b \bigm\vert \msf{Y}^{b-1}, \msf{Y}_b\notin\mc{Y}^{b-1} \bigr) 
    + \Pp(\msf{Y}_b\in\mc{Y}^{b-1}) H\bigl(\msf{Y}_b \bigm\vert \msf{Y}^{b-1}, \msf{Y}_b\in\mc{Y}^{b-1} \bigr).
\end{align}

Conditioned on $\msf{Y}_b\notin\mc{Y}^{b-1}$, $\msf{Y}^{b-1}$, and $\msf{L}_b$, the
source block $\msf{Y}_b$ is uniformly distributed over
$\{0,1\}^{\msf{L}_b}\setminus\mc{Y}^{b-1}$. Hence,
\begin{align*}
    H\bigl(\msf{Y}_b \bigm\vert \msf{Y}^{b-1}, \msf{Y}_b\notin\mc{Y}^{b-1} \bigr)
    & \geq H\bigl(\msf{Y}_b \bigm\vert \msf{Y}^{b-1}, \msf{Y}_b\notin\mc{Y}^{b-1}, \msf{L}_b \bigr) \\
    & = \E\bigl( \log\card{\{0,1\}^{\msf{L}_b}\setminus\mc{Y}^{b-1}} \bigm\vert \msf{Y}_b\notin\mc{Y}^{b-1} \bigr) \\
    & \geq \E\bigl( \log(2^{\msf{L}_b}-\card{\mc{Y}^{b-1}}) \bigm\vert \msf{Y}_b\notin\mc{Y}^{b-1} \bigr) \\
    & \geq \E\log(2^{\msf{L}_b}-A),
\end{align*}
where we have used the independence of $\msf{L}_b$ and the event
$\msf{Y}_b\notin\mc{Y}^{b-1}$. Using the assumption that $A \leq
2^{\E(\msf{L}_b)/2-1} \leq 2^{\msf{L}_b-1}$, this last expression can be further
lower bounded as
\begin{equation*}
    \E\log(2^{\msf{L}_b}-A) 
    \geq \E(\msf{L}_b)-1
    = \E(\msf{L})-1,
\end{equation*}
so that
\begin{equation}
    \label{eq:variabl_lower5}
    \Pp(\msf{Y}_b\notin\mc{Y}^{b-1})H\bigl(\msf{Y}_b \bigm\vert \msf{Y}^{b-1}, \msf{Y}_b\notin\mc{Y}^{b-1} \bigr)
    \geq \Pp(\msf{Y}_b\notin\mc{Y}^{b-1})(\E(\msf{L})-1).
\end{equation}
Furthermore, by the same arguments as in~\eqref{eq:fixed_lower5} in
Appendix~\ref{sec:proofs_fixed},
\begin{align}
    \label{eq:variable_lower6}
    \Pp(\msf{Y}_b\in\mc{Y}^{b-1}) H\bigl(\msf{Y}_b \bigm\vert \msf{Y}^{b-1}, \msf{Y}_b\in\mc{Y}^{b-1} \bigr) 
    & = A^{-1}\E\bigl( \card{\mc{Y}^{b-1}} \log\card{\mc{Y}^{b-1}} \bigr) \notag\\
    & \geq A^{-1}\E\card{\mc{Y}^{b-1}} \log\E\card{\mc{Y}^{b-1}},
\end{align}
where the last line follows from Jensen's inequality.

Combining~\eqref{eq:variable_lower1}--\eqref{eq:variable_lower6} yields
\begin{equation}
    \label{eq:variable_lower7}
    \Rs 
    \geq (\E(\msf{L})-1)\sum_{b=1}^B \Pp(\msf{Y}_b\notin\mc{Y}^{b-1})
    + A^{-1}\sum_{b=1}^B\E\card{\mc{Y}^{b-1}}\log\E\card{\mc{Y}^{b-1}}
    - B(1+\log\E(\msf{L})).
\end{equation}
To obtain a more explicit expression, we can further lower bound $\Rs$ as
\begin{align}
    \label{eq:variable_lower8}
    \Rs \geq \bigl( 0.5(\E(\msf{L})-1)\min\{A,B\} + 0.5(B-A)^+\log(A/2) - B(1+\log\E(\msf{L})) \bigr)^+,
\end{align}
similar to~\eqref{eq:fixed_lower7} in Appendix~\ref{sec:proofs_fixed}.

From~\eqref{eq:variable_upper13} and~\eqref{eq:variable_lower7}, we obtain
\begin{align}
    \label{eq:variable1}
    \Rv-\Rs
    & \leq 2\log\bigl(B\E(\msf{L})\bigr) +3
    +B\bigl(4+2^{1-M}\E(\msf{L})+\log\E(\msf{L})\bigr) \notag\\
    & \quad {} + B2^{M+2}  
    +B\bigl(1+2^{-M}\E(\msf{L})\bigr) \log\bigl( 2B\E(\msf{L}) \bigr).
\end{align}
The two dominant terms in this last expression behave (to first order) like
$2^MB$ and $2^{-M}B\E(\msf{L})$. Hence, the right-hand side
of~\eqref{eq:variable1} is approximately minimized by choosing the anchor length
as
\begin{equation*}
    M \defeq \ceil{0.5\log\E(\msf{L})}. 
\end{equation*}
This splits each source
block into an average of about $\sqrt{\E(\msf{L})}$ chunks of expected length
about $\sqrt{\E(\msf{L})}$. With this choice of $M$, \eqref{eq:variable1} yields
\begin{align*}
    \Rv-\Rs
    & \leq 2\log\bigl(B\E(\msf{L})\bigr) + 3
    +B\bigl(4+2\sqrt{\E(\msf{L})}+\log\E(\msf{L})\bigr) \notag\\ 
    & \quad {} +  8B\sqrt{\E(\msf{L})} +B\bigl(1+\sqrt{\E(\msf{L})}\bigr)
    \log\bigl( 2B\E(\msf{L}) \bigr) \notag\\
    & \leq 2B(1+\sqrt{\E(\msf{L})}) \log\bigl( B\E(\msf{L}) \bigr),
\end{align*}
where the last inequality holds for $B$ large enough.  Combining this
with~\eqref{eq:variable_lower8} yields
\begin{align*}
    \frac{\Rv}{\Rs}-1
    & \leq \frac{4B(1+\sqrt{\E(\msf{L})}) \log\bigl( B\E(\msf{L}) \bigr)}
    {\bigl((\E(\msf{L})-1)\min\{A,B\} + (B-A)^+\log(A/2) - 2B\log(2\E(\msf{L}))\bigr)^+},
\end{align*}
again for $B$ large enough. This proves the theorem.
\hfill\IEEEQED

\section{Analysis of Variable-Length Deduplication for $B > A^2$ (Example~\ref{eg:sync})}
\label{sec:proofs_sync2}

The bound~\eqref{eq:variable_upper11} in Appendix~\ref{sec:proofs_variable} is
appropriate when $B \leq A^2$. When $B > A^2$, it can be quite loose, since each
pair $\msf{X}_a\msf{X}_{\tilde{a}}$ yields at most three distinct boundary
deduplication chunks. We next derive a tighter bound for the regime $B > A^2$.

Define the event $\mc{E}$ that at least one source symbol $\msf{X}\in\mc{X}$
does not contain the substring $10^M$. Then, since each pair
$\msf{X}_a\msf{X}_{\tilde{a}}$ yields at most three distinct boundary
deduplication chunks, we have on the complement of
$\mc{E}$ that
\begin{align*}
    \sum_{b=1}^B\sum_{c\in\partial\mc{C}_b}\ind_{\{\msf{Z}_c\notin\mc{Z}^{c-1}\}} \ell(\msf{Z}_c) 
    & \leq \sum_{a=1}^A\sum_{\tilde{a}=1}^A
    \bigl(\ell(\tail(\msf{X}_a)) + \ell(\head(\msf{X}_{\tilde{a}})) \bigr) \\
    & = A\sum_{a=1}^A\bigl(\ell(\tail(\msf{X}_a)) + \ell(\head(\msf{X}_a)) \bigr),
\end{align*}
where $\head(\msf{X}_a)$ is the string starting from the beginning of
$\msf{X}_a$ up to and including the first occurrence of $10^M$, and where
$\tail(\msf{X}_a)$ is the string from the end of $\msf{X}_a$ backwards until the
end of the first (counting backwards) occurrence of $0^M$ (see
Fig.~\ref{fig:boundary} in Appendix~\ref{sec:proofs_variable}). On $\mc{E}$, we
have
\begin{equation*}
    \sum_{b=1}^B\sum_{c\in\partial\mc{C}_b}\ind_{\{\msf{Z}_c\notin\mc{Z}^{c-1}\}} \ell(\msf{Z}_c)
    \leq \ell(\msf{S}) \leq 2B\E(\msf{L}).
\end{equation*}
Combining these last two inequalities yields
\begin{equation*}
     \E\biggl(\sum_{b=1}^B\sum_{c\in\partial\mc{C}_b}\ind_{\{\msf{Z}_c\notin\mc{Z}^{c-1}\}}\ell(\msf{Z}_c)\biggr)
     \leq A\sum_{a=1}^A\E\bigl(\ell(\tail(\msf{X}_a)) + \ell(\head(\msf{X}_a)) \bigr)
     + 2B\E(\msf{L})\Pp(\mc{E}).
\end{equation*}

We have
\begin{align*}
    A\sum_{a=1}^A\E\bigl(\ell(\tail(\msf{X}_a)) + \ell(\head(\msf{X}_a)) \bigr)
    & \leq A^2 2^{M+2},
\end{align*}
where we have again used \cite[Theorems~8.2~and~8.3]{sedgewick13}.  It remains to analyze
$\Pp(\mc{E})$. Since $\msf{L} \geq \E(\msf{L})/2$ by assumption, the
probability of the event $\mc{E}$ is upper bounded by that of the event that
from $A$ sequences drawn uniformly at random without replacement from
$\{0,1\}^{\E(\msf{L})/2}$ at least one of them contains no occurrence of $10^M$.
The probability of this last event is upper bounded as
\begin{align*}
    \Pp(\mc{E}) 
    & \leq A\bigl(1-2^{-M-1}\bigr)^{\floor{\E(\msf{L})/(2M+2)}} \\
    & \leq A\exp\bigl(-2^{-M-1}\floor{\E(\msf{L})/(2M+2)})\bigr).
\end{align*}
Hence,
\begin{equation}
    \label{eq:variable_upper11_alt} 
     \E\biggl(\sum_{b=1}^B\sum_{c\in\partial\mc{C}_b}\ind_{\{\msf{Z}_c\notin\mc{Z}^{c-1}\}}\ell(\msf{Z}_c)\biggr)
     \leq A^2 2^{M+2} + 2AB\E(\msf{L})\exp\bigl(-2^{-M-1}\floor{\E(\msf{L})/(2M+2)})\bigr).
\end{equation}

Substituting~\eqref{eq:variable_upper7}--\eqref{eq:variable_upper9},
\eqref{eq:variable_upper10}, and~\eqref{eq:variable_upper11_alt} into
\eqref{eq:variable_upper6} in Appendix~\ref{sec:proofs_variable} yields
\begin{align*}
    \Rv
    & \leq 
    2\log\bigl(B\E(\msf{L})\bigr)
    +3
    +2B\bigl(1+2^{-M}\E(\msf{L})\bigr)
    +\E(\msf{L})\sum_{b=1}^B \Pp(\msf{Y}_b\notin\mc{Y}^{b-1}) \notag\\
    & \quad {} +B\bigl(1+2^{-M}\E(\msf{L})\bigr) \log\bigl( 2B\E(\msf{L}) \bigr)
    + A^2 2^{M+2} \notag\\
    & \quad {} + 2AB\E(\msf{L})\exp\bigl(-2^{-M-1}\floor{\E(\msf{L})/(2M+2)})\bigr).
\end{align*}
Combined with~\eqref{eq:variable_lower7}, this shows that
\begin{align}
    \label{eq:variable_upper12_alt} 
    \Rv-\Rs
    & \leq 
    2\log\bigl(B\E(\msf{L})\bigr)
    +3
    +B\bigl(4+2^{1-M}\E(\msf{L})+\log\E(\msf{L})\bigr) \notag\\
    & \quad {} +B\bigl(1+2^{-M}\E(\msf{L})\bigr) \log\bigl( 2B\E(\msf{L}) \bigr)
    + A^2 2^{M+2} \notag\\
    & \quad {} + 2AB\E(\msf{L})\exp\bigl(-2^{-M-1}\floor{\E(\msf{L})/(2M+2)})\bigr).
\end{align}

For the remainder of the argument, we specialize to the setting in
Example~\ref{eg:sync}, namely $A=2$, $B=3L$, $\Pp(\msf{L} = L) = \Pp(\msf{L} =
L+1) = 1/2$. With this, \eqref{eq:variable_upper12_alt} becomes
\begin{equation*}
    \Rv-\Rs \leq O\Bigl(
    L(1+2^{-M}L)\log L + 2^M + L^2\exp\bigl(-2^{-M-2}L/(M+1)\bigr)
    \Bigr).
\end{equation*}
Set the anchor length to
\begin{equation*}
    M \defeq \floor{\log L - 2\log\log L-2},
\end{equation*}
which results in
\begin{equation}
    \label{eq:variable_upper13_alt} 
    \Rv-\Rs \leq O( L\log^3 L ).
\end{equation}

Now, since there are only $A=2$ source symbols of length $L$ or $L+1$, we can
with high probability uniquely identify the source blocks $\msf{Y}_1, \msf{Y}_2,
\dots, \msf{Y}_B$ from $\msf{S}$ for $L$ large enough. Hence, each source block
$\msf{Y}_b$ adds asymptotically one bit of information to $\msf{S}$, i.e., 
\begin{equation*}
    \Rs \geq H(\msf{S}) \geq (1-o(1)) B = (1-o(1)) 3L
\end{equation*}
as $L\to\infty$.
Combining this with~\eqref{eq:variable_upper13_alt} shows that
\begin{equation*}
    \frac{\Rv}{\Rs} \leq O(\log^3L) = O(\log^3B),
\end{equation*}
as claimed. \hfill\IEEEQED

\section{Analysis of Example~\ref{eg:duplicate}}
\label{sec:proofs_duplicate}

Recall that $A = \sqrt{B}$ and $\msf{L} = L = \sqrt{B}$. Following the same
steps as those leading to~\eqref{eq:fixed2_upper1} in
Appendix~\ref{sec:proofs_sync}, we obtain the upper bound
\begin{align}
    \label{eq:duplicate1}
    \Rs 
    & \leq AL+B\log(A)+1 \notag\\
    & = B+0.5B\log(B)+1 \notag\\
    & \leq O\bigl(B\log(B)\bigr)
\end{align}
for the rate $\Rs$ of the optimal source code.

We continue with a lower bound on the rate $\Rv$ of variable-length
deduplication. For each chunk $c$, the flag indicating if the chunk is already
in the dictionary takes one bit, resulting in a total of $\E(\msf{C})$ bits.
Further, each unique boundary chunk needs to be stored in the dictionary. We
next argue that we will see on the order of $A^2 = B$ unique boundary chunks
that each have length on the order of $\min\{2^M,L\}$. This will imply that
storing the unique boundary chunks takes on the order of $\min\{2^M,L\}B$ bits. 

\begin{figure}[htbp]
    \centering
    \includegraphics{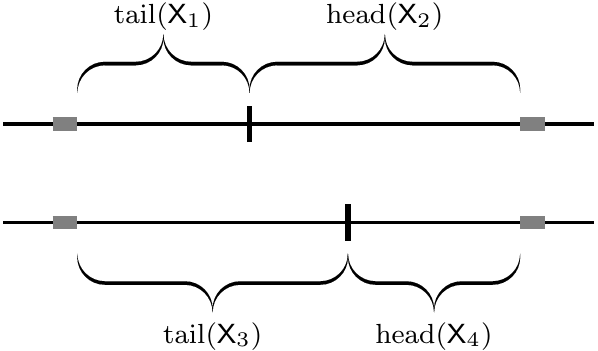}
    \caption{Duplicate boundary deduplication chunks arising from the concatenation of
        different source symbols $\msf{X}_1\msf{X}_2$ and $\msf{X}_3\msf{X}_4$.
        Compare to Fig.~\ref{fig:anchors} in Section~\ref{sec:results}.}
    \label{fig:duplicate}
\end{figure} 

Consider the two concatenations of source symbols $\msf{X}_1\msf{X}_2$ and
$\msf{X}_3\msf{X}_4$, and consider the two resulting boundary deduplication
chunks (see Fig.~\ref{fig:duplicate}). We can decompose the two boundary chunks
as the concatenation $\tail(\msf{X}_1)\head(\msf{X}_2)$ and
$\tail(\msf{X}_3)\head(\msf{X}_4)$, where $\tail(\cdot)$ and $\head(\cdot)$
denote the substring of the source symbol contributing to the boundary chunk
(excluding the anchor sequence, and truncated to length $L/2$ in case there is
no anchor sequence before then). From Fig.~\ref{fig:duplicate} we see that if
$\tail(\msf{X}_1)\head(\msf{X}_2) = \tail(\msf{X}_3)\head(\msf{X}_4)$, then one
of $\head(\msf{X}_2)$, $\head(\msf{X}_4)$ is a substring of the other, and one of
$\tail(\msf{X}_1)$, $\tail(\msf{X}_3)$ is a substring of the other. 

Consider a source symbol $\msf{X}_a$, and assume $M\geq 10$ for now. With
probability at least
\begin{equation*}
    1-2\cdot2^{M-5}\cdot2^{-M}-2\cdot2^{-M/2} 
    \geq 1-2^{-3}
\end{equation*}
it has $\head(\msf{X}_a)$ of length at least $\min\{2^{M-5}, L/2\}$ and containing a
least one symbol $1$ in the first  $M/2$ bits, and it has $\tail(\msf{X}_a)$ of
length at least $\min\{2^{M-5}, L/2\}$ and containing a least one symbol $1$ in
the last $M/2$ bits. A short calculation (using Markov's inequality) shows that
this implies that with probability at least $3/4$, the source alphabet $\mc{X}$
has at least $A/8$ symbols with this property.

Moreover, with probability at least $1-(AL)^22^{-2^{M-5}}$ the source alphabet
$\mc{X}$ has no repeating, nonoverlapping substrings of size $2^{M-5}$. This
argument is reported with more detail in Appendix~\ref{sec:proofs_multi}. In
particular, if 
\begin{equation}
    \label{eq:duplicate_mcond}
    M 
    \geq 5+\log\log\bigl(8(AL)^2\bigr) 
    = 5+\log\log\bigl(8B^2\bigr),
\end{equation}
then with probability at least $3/4$ the source alphabet $\mc{X}$ has no
repeating, nonoverlapping substrings of size $\min\{2^{M-5}, L/2\}$.

Combining the two arguments shows that with probability at least $1/2$ there
are at least $A/8$ source symbols with both long, duplicate-free heads and
tails. Further, each of these heads contains a symbol one within the first $M/2$
bits, and each of these tails contains a symbol one within its first $M/2$ bits.
If this event holds, then $A^2/64 = B/64$ of all $A^2$ possible concatenations
$\msf{X}_a\msf{X}_{\tilde{a}}$ produce unique boundary chunks of length at least
$\min\{2^{M-4}, L\}$. 

Finally, since $B = A^2$, we will see at least $1/2$ of these possible unique
boundary chunks in expectation. Therefore, the expected number of bits needed to
store just the boundary chunks is at least
$\Omega\bigl(\min\{2^M,L\}B\bigr)\geq\Omega\bigl(\min\{2^M,B^{1/2}\}B\bigr)$,
assuming~\eqref{eq:duplicate_mcond} is satisfied.

Combining these arguments, the rate $\Rv$ of variable-length deduplication is
lower bounded as
\begin{equation*}
    \Rv \geq \E(\msf{C})+\ind_{\{M\geq 5+\log\log(8B^2)\}}\Omega\bigl(\min\{2^M,B^{1/2}\}B\bigr),
\end{equation*}
The expected number of chunks $\E(\msf{C})$ is lower bounded by $2^{-M}BL/M = 2^{-M}B^{3/2}/M$, so that
\begin{equation*}
    \Rv \geq 2^{-M}B^{3/2}/M+\ind_{\{M\geq 5+\log\log(8B^2)\}}\Omega\bigl(\min\{2^M,B^{1/2}\}B/M\bigr).
\end{equation*}
This lower bound is minimized by
\begin{equation*}
    M \defeq \tfrac{1}{4}\log(B) + O(1),
\end{equation*}
which results in the bound
\begin{equation}
    \label{eq:duplicate2}
    \Rv \geq \Omega(B^{5/4}\log^{-1}(B)) 
\end{equation}
as $B\to\infty$.

Combining~\eqref{eq:duplicate1} and~\eqref{eq:duplicate2} yields
\begin{equation*}
    \frac{\Rv}{\Rs} 
    \geq \Omega\bigl( B^{1/4}\log^{-2}(B) \bigr)
\end{equation*}
as $B\to\infty$. \hfill\IEEEQED

\section{Analysis of Multi-Chunk Deduplication (Proof of Theorem~\ref{thm:multi})}
\label{sec:proofs_multi}

We start with an upper bound on the rate $\Rm$ of multi-chunk deduplication.
The initial encoding of the length $\ell(\msf{S})$ using a universal code for
the integers takes again at most $2\log\ell(\msf{S})+3$ bits by
\cite[Lemma~13.5.1]{cover06}. Consider then the encoding of chunk $c$. The flag
indicating if the chunk is already in the dictionary takes one bit. If the chunk
is new, then $\msf{V}_c$ is encoded using at most $2\log(\msf{V}_c)+3$ bits,
plus the $\msf{V}_c$ chunks starting at $c$ are added to the dictionary using
$\ell\bigl(\msf{Z}_c\msf{Z}_{c+1}\cdots\msf{Z}_{c+\msf{V}_c-1}\bigr)$ bits. If
the chunk is already in the dictionary, then $\msf{W}_c$ is encoded using at
most $2\log(\msf{W}_c)+3$ bits, plus a pointer into the dictionary using at most
$\log\card{\mc{Z}^{c-1}}+1$ bits, where $\mc{Z}^{c-1}$ is again the dictionary
when encoding chunk $c$. The next chunk to be encoded is either $c+\msf{V}_c$ or
$c+\msf{W}_c$.

\begin{figure}[htbp]
    \centering
    \includegraphics{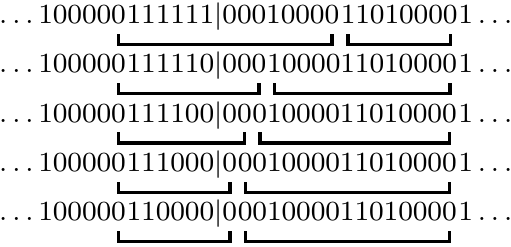}
    \caption{Boundary chunks (indicated by underbraces) as defined by
        multi-chunk deduplication for anchor length $M=4$.
        The vertical line $\vert$ indicates the boundary between the source
        blocks $\msf{Y}_{b-1}$ and $\msf{Y}_b$.}
    \label{fig:boundary_mult}
\end{figure} 

As before, we denote by $\mc{C}_b$ those chunks starting in source block
$\msf{Y}_b$. We again define the notion of boundary chunk indices
$\partial\mc{C}_b$ and interior chunk indices $\mc{C}_b^\circ$ (see
Appendix~\ref{sec:proofs_variable}), but this time with respect to the
multi-chunk deduplication scheme, as shown in Fig.~\ref{fig:boundary_mult}.

Let $\mc{E}$ be the event that there is at least one interior chunk of
$\msf{X}_1, \msf{X}_2, \dots, \msf{X}_A$ that is either equal to another
interior chunk in the source alphabet or to a boundary chunk of $\msf{S}$.
Assume we are on the complement of $\mc{E}$ for now, and consider the first
interior chunk $\msf{Z}_c$ in $\msf{Y}_b$ (i.e., the smallest
$c\in\mc{C}_b^\circ$). Then there are two possibilities, either
$\msf{Y}_b\notin\mc{Y}^{b-1}$ and all indices $\mc{C}_b^\circ$ correspond to new
chunks, or $\msf{Y}_b\in\mc{Y}^{b-1}$ and all indices $\mc{C}_b^\circ$
correspond to chunks already in the dictionary. 

In the former case ($\msf{Y}_b\notin\mc{Y}^{b-1}$), the number $\msf{V}_c$ is
larger than or equal to $\card{\mc{C}_b^\circ}$. The encoding of the chunks with
indices in $\mc{C}_b^\circ$ takes thus at most
\begin{equation*}
    4+2\log(\msf{V}_c)+\ell\bigl(\msf{Z}_c\msf{Z}_{c+1}\cdots\msf{Z}_{c+\msf{V}_c-1}\bigr)
\end{equation*}
bits. Since reducing the number of jointly encoded chunks and encoding them
separately increases the aggregate rate, we can upper bound the total
rate by assuming that $\msf{V}_c = \card{\mc{C}_b^\circ}$. The encoding for the
first interior chunk $c$ in $\mc{C}_b^\circ$ takes in this case at most
\begin{equation*}
    4+2\log\card{\mc{C}_b^\circ}+\ell(\msf{Y}_b)
\end{equation*}
bits.

In the latter case ($\msf{Y}_b\in\mc{Y}^{b-1}$), the number $\msf{W}_c$ is
larger than or equal to $\card{\mc{C}_b^\circ}$ (since the corresponding chunks
must have been seen in sequence by the uniqueness assumption), and all the
chunks with indices in $\mc{C}_b^\circ$ are encoded together, taking at most
\begin{equation*}
    5+2\log(\msf{W}_c)+\log\card{\mc{Z}^{c-1}}
\end{equation*}
bits. Again, reducing the number of jointly encoded chunks and encoding them
separately increases the aggregate rate, and thus we can upper bound the total
rate by assuming that $\msf{W}_c = \card{\mc{C}_b^\circ}$.  The encoding for the
first interior chunk $c$ in $\mc{C}_b^\circ$ takes in this case at most
\begin{equation*}
    5+2\log\card{\mc{C}_b^\circ}+\log\card{\mc{Z}^{c-1}}
\end{equation*}
bits.

Assume next that we are on $\mc{E}$. If an interior chunk $\msf{Z}_c$ is not in
the dictionary, it is encoded in the worst case using $4+\ell(\msf{Z}_c)$ bits. If
an interior chunk $\msf{Z}_c$ is in the dictionary, then the pointer into the
dictionary takes at most $1+\log\bigl(2B\E(\msf{L})\bigr)$ bits. In the worst
case, each old chunk is encoded separately, leading to an additional $3$ bits
for the encoding of $\msf{W}_c$. Thus, as long as 
\begin{equation}
    \label{eq:mult_mcond}
    \log\bigl(2B\E(\msf{L})\bigr) \leq 2^{M-1}
\end{equation}
the encoding of each old interior chunk $\msf{Z}_c$ takes at most 
\begin{equation*}
    5+\log\bigl(2B\E(\msf{L})\bigr)
    \leq 5+2^{M-1} 
    \leq 5+\ell(\msf{Z}_c)
\end{equation*}
bits, since each chunk has length at least $2^{M-1}$ by
construction. Thus, on the complement of $\mc{E}$ and
assuming~\eqref{eq:mult_mcond} is satisfied, the encoding of the interior chunks
of $\msf{S}$ takes at most 
\begin{equation*}
    \sum_{c=1}^{\msf{C}} \bigl(5+\ell(\msf{Z}_c)\bigr)
    \leq 6\ell(\msf{S}) 
    \leq 12B\E(\msf{L})
\end{equation*}
bits.

Similarly, as long as the condition~\eqref{eq:mult_mcond} is satisfied, the
boundary chunks can be encoded using at most $5+\ell(\msf{Z}_c)$
bits each, regardless of whether they are in the dictionary.

We can then upper bound the rate $\Rm(\msf{S})$ for a particular source sequence
$\msf{S}$ as
\begin{align*}
    \Rm(\msf{S})
    & \leq 2\log\ell(\msf{S})+3
    + \sum_{b=1}^B\ind_{\{\msf{Y}_b\notin\mc{Y}^{b-1}\}}\bigl( 4+2\log\card{\mc{C}_b^\circ}+\ell(\msf{Y}_b) \bigr) \notag\\
    & \quad {} + \sum_{b=1}^B\ind_{\{\msf{Y}_b\in\mc{Y}^{b-1}\}}\bigl( 5+2\log\card{\mc{C}_b^\circ}+\log\card{\mc{Z}^{c-1}} \bigr)
    + \sum_{b=1}^B\sum_{c\in\partial\mc{C}_b}\bigl(5+\ell(\msf{Z}_c)\bigr)
    + \ind_{\mc{E}}12B\E(\msf{L}) \notag\\
    & \leq 5+2\log\bigl(B\E(\msf{L})\bigr)+B\bigl(8+\log(B)+3\log\E(\msf{L}) \bigr) 
    + \sum_{b=1}^B\ind_{\{\msf{Y}_b\notin\mc{Y}^{b-1}\}}\ell(\msf{Y}_b) \notag\\
    & \quad {} + \sum_{b=1}^B\sum_{c\in\partial\mc{C}_b}\bigl(5+\ell(\msf{Z}_c)\bigr)
    + \ind_{\mc{E}}12B\E(\msf{L}),
\end{align*}
where the second inequality follows after some algebra using the bounds
$\card{\mc{C}_b^\circ} \leq 2\E(\msf{L})$, $\card{\mc{Z}^{c-1}} \leq
2B\E(\msf{L})$, and $\ell(\msf{S}) \leq 2B\E(\msf{L})$.
Taking expectations yields
\begin{align}
    \label{eq:mult_upper1}
    \Rm
    & \leq 5+2\log\bigl(B\E(\msf{L})\bigr)+B\bigl(8+\log(B)+3\log\E(\msf{L}) \bigr) 
    + \E(\msf{L})\sum_{b=1}^B\Pp(\msf{Y}_b\notin\mc{Y}^{b-1}) \notag\\
    & \quad {} + \sum_{b=1}^B\E\biggl(\sum_{c\in\partial\mc{C}_b}\bigl(5+\ell(\msf{Z}_c)\bigr)\biggr)
    + 12B\E(\msf{L})\Pp(\mc{E}),
\end{align}
It remains to upper bound the last two terms in~\eqref{eq:mult_upper1}.

For the second-to-last term in~\eqref{eq:mult_upper1}, we have the upper bound
\begin{equation}
    \label{eq:mult_upper2}
    \sum_{b=1}^B
    \E\biggl(\sum_{c\in\partial\mc{C}_b}\bigl(5+\ell(\msf{Z}_c)\bigr)\biggr)
    \leq \sum_{b=1}^B
    \Bigl( 5\E\card{\partial\mc{C}_b} + \E \ell\bigl(\tail(\msf{Y}_b)\bigr)
    +\E \ell\bigl(\head(\msf{Y}_b)\bigr) \Bigr).
\end{equation}
Here, $\head(\msf{Y}_b)$ are the bits from the start of $\msf{Y}_b$ forward
until the end of the first (counting forwards) occurrence of the string $u10^M$
such that $u\in\{0,1\}^{2^{M-1}-M-1}$ does not contain $0^M$. And
$\tail(\msf{Y}_b)$ are the bits from the end of $\msf{Y}_b$ backwards until the
end of the first (counting backwards) occurrence of the string $0^M$ plus an
additional $2^{M-1}-M$ bits. If no such substring occurs, head and tail denote
all the bits in $\msf{Y}_b$ in either case.

The expected value of $\ell\bigl(\tail(\msf{Y}_b)\bigr)$ is upper bounded as
\begin{equation}
    \label{eq:mult_upper3a}
    \E\ell\bigl(\tail(\msf{Y}_b)\bigr) \leq 2^{M+1}-2+2^{M-1}-M \leq 2^{M+2}
\end{equation}
as before by \cite[Theorem~8.2]{sedgewick13} (see again
Appendix~\ref{sec:proofs_variable}). The quantity
$\E\ell\bigl(\head(\msf{Y}_b)\bigr)$ can be upper bounded by replacing
$\msf{Y}_b$ with an infinite-length $\Bernoulli(1/2)$ process. The head of that
Bernoulli process is then a concatenation of sub-chunks of the form $\msf{U}_1 0^M \msf{U}_2
0^M\cdots \msf{U}_\msf{N} 0^M$ with  $\ell(\msf{U}_n) < 2^{M-1}-M$ for $n < \msf{N}$ and
with $\ell(\msf{U}_\msf{N}) \geq 2^{M-1}-M$. Consider the sequence of lengths
$\ell(\msf{U}_1)$, $\ell(\msf{U}_2)$, \dots. This sequence forms an \iid stochastic
process, and $\msf{N}$ is a stopping time with respect to this process. Thus, we
can apply Wald's equation together with \cite[Theorem~8.2]{sedgewick13} to
obtain
\begin{equation*}
    \E\ell\bigl(\head(\msf{Y}_b)\bigr) \leq \E(\msf{N}) 2^{M+1}.
\end{equation*}
The random variable $\msf{N}$ is geometrically distributed with probability of
success lower bounded by $1-2^{M-1}\cdot 2^{-M} = 1/2$. Thus $\E(\msf{N}) \leq
2$ and
\begin{equation}
    \label{eq:mult_upper3b}
    \E\ell\bigl(\head(\msf{Y}_b)\bigr) \leq 2^{M+2}.
\end{equation}
The same argument also shows that
\begin{equation}
    \label{eq:mult_upper3c}
   \E\card{\partial\mc{C}_b} \leq \E(\msf{N})+1 \leq 3.
\end{equation}
Substituting \eqref{eq:mult_upper3a}--\eqref{eq:mult_upper3c}
into~\eqref{eq:mult_upper2}, we obtain
\begin{equation}
    \label{eq:mult_upper3}
    \sum_{b=1}^B
    \E\biggl(\sum_{c\in\partial\mc{C}_b}\bigl(5+\ell(\msf{Z}_c)\bigr)\biggr)
    \leq B\bigl(15+2^{M+3}\bigr).
\end{equation}

For the last term $12B\E(\msf{L})\Pp(\mc{E})$ in~\eqref{eq:mult_upper1}, we need
to upper bound the probability that there is at least one interior chunk in the
source alphabet that is either equal to another interior chunk of the source
alphabet or to a boundary chunk of the source sequence. Since all chunks have
length at least $2^{M-1}$ by construction, whenever this last event holds, then
$\msf{X}_1, \msf{X}_2, \dots, \msf{X}_A$ contains a nonoverlapping duplicate
substring of length $2^{M-2}$ (where the additional factor $1/2$ accounts for
the boundary chunks). 

Consider a source symbol, say $\msf{X}_1$. The probability that it contains a
nonoverlapping duplicate substring of length $2^{M-2}$ is upper bounded by the
probability that a $\Bernoulli(1/2)$ process of length $2\E(\msf{L})$ contains
such a substring. Consider next two source symbols, say $\msf{X}_1$ and
$\msf{X}_2$. Condition on their lengths $\ell(\msf{X}_1)$ and $\ell(\msf{X}_2)$.
If these lengths are distinct, then $\msf{X}_1$ and $\msf{X}_2$ are independent
$\Bernoulli(1/2)$ processes of given length. If the lengths are the same, then
$\msf{X}_1$ and $\msf{X}_2$ are not independent, since they are chosen without
replacement. However, the probability of $\msf{X}_1$ containing a duplicate
substring of length $2^{M-2}$ from $\msf{X}_2$ is upper bounded by drawing them
with replacement. Further, in both cases, the probability of the event under
consideration is increased if we increase the length of the source symbols.  In
summary, the probability that the source alphabet $\mc{X}$ contains a
nonoverlapping duplicate substring of length $2^{M-2}$ is upper bounded by the
probability that $A$ independent $\Bernoulli(1/2)$ processes of length
$2\E(\msf{L})$ contain such a duplicate. This probability can in turn be upper
bounded by $\bigl(2A\E(\msf{L})\bigr)^22^{-2^{M-2}}$. Therefore,
\begin{equation}
    \label{eq:mult_upper4}
    12B\E(\msf{L})\Pp(\mc{E}) 
    \leq 48A^2B\E(\msf{L})^32^{-2^{M-2}}.
\end{equation}

Substituting~\eqref{eq:mult_upper3} and~\eqref{eq:mult_upper4}
into~\eqref{eq:mult_upper1} yields
\begin{align*}
    \label{eq:mult_upper1}
    \Rm
    & \leq 5+2\log\bigl(B\E(\msf{L})\bigr)+B\bigl(23+\log(B)+3\log\E(\msf{L})+2^{M+3}\bigr) \notag\\
    & \quad {} + \E(\msf{L})\sum_{b=1}^B\Pp(\msf{Y}_b\notin\mc{Y}^{b-1})
    + 48A^2B\E(\msf{L})^32^{-2^{M-2}}.
\end{align*}
Combined with~\eqref{eq:variable_lower7} in Appendix~\ref{sec:proofs_variable},
this shows that
\begin{equation}
    \label{eq:mult_upper5}
    \Rm-\Rs
    \leq 5+2\log\bigl(B\E(\msf{L})\bigr)+B\bigl(25+\log(B)+4\log\E(\msf{L})+2^{M+3}\bigr)
    + 48A^2B\E(\msf{L})^32^{-2^{M-2}}.
\end{equation}

Assume first that $A^4\E(\msf{L})^5\geq 2B$, and set
\begin{equation*}
    M \defeq \ceil{\log\log\bigl(A^2\E(\msf{L})^3\bigr)+2}.
\end{equation*}
Note that
\begin{equation*}
    2^{M-1} 
    \geq \log\bigl(A^4 \E(\msf{L})^6\bigr)
    \geq \log\bigl(2B \E(\msf{L})\bigr),
\end{equation*}
satisfying \eqref{eq:mult_mcond}. With this
choice of $M$, \eqref{eq:mult_upper5} becomes after some simplification
\begin{equation}
    \label{eq:mult_upper6a}
    \Rm-\Rs
    \leq 5+2\log\bigl(B\E(\msf{L})\bigr)+B\bigl(73+128\log(A)+\log(B)+196\log\E(\msf{L})\bigr).
\end{equation}
Assume next that $A^4\E(\msf{L})^5 < 2B$, and set
\begin{equation*}
    M \defeq \ceil{\log\log\bigl(2B\E(\msf{L})\bigr)+1}.
\end{equation*}
Note that then
\begin{equation*}
    2^{M-1} \geq \log\bigl(2B \E(\msf{L})\bigr),
\end{equation*}
again satisfying \eqref{eq:mult_mcond}. With this choice of $M$,
\eqref{eq:mult_upper5} becomes after some simplification
\begin{equation}
    \label{eq:mult_upper6b}
    \Rm-\Rs
    \leq 5+2\log\bigl(B\E(\msf{L})\bigr)+B\bigl(105+33\log(B)+36\log\E(\msf{L})\bigr).
\end{equation}

From~\eqref{eq:mult_upper6a} and~\eqref{eq:mult_upper6b}, we conclude that,
regardless of the relationship of $A$, $B$, and $\E(\msf{L})$, we have
\begin{equation*}
    \label{eq:mult_upper6}
    \Rm-\Rs
    \leq O\bigl(B\log\bigl(AB\E(\msf{L})\bigr)\bigr)
\end{equation*}
as $B\to\infty$.  Together with~\eqref{eq:variable_lower8} in
Appendix~\ref{sec:proofs_variable}, we obtain
\begin{align}
    \frac{\Rm}{\Rs}-1 
    \leq
    O\biggl(\frac{B\log\bigl(AB\E(\msf{L})\bigr)}
    {\bigl((\E(\msf{L})-1)\min\{A,B\} + (B-A)^+\log(A/2) - 2B\log(2\E(\msf{L}))\bigr)^+}\biggr)
\end{align}
as $B\to\infty$.\hfill\IEEEQED

\section*{Acknowledgments}

The author thanks M. A. Maddah-Ali for helpful initial discussions and
the reviewers for their comments.

\end{document}